\documentclass[a4paper,USenglish, cleveref, autoref, thm-restate, numberwithinsect]{lipics-v2021}
\usepackage[usenames,dvipsnames]{xcolor} 
 
\usepackage[dvipsnames]{xcolor}
\usepackage{soul}

\usepackage{algorithm}  
\usepackage{algorithmicx}  
\usepackage[noend]{algpseudocode}
\usepackage{amsmath}

\algnewcommand\And{\textbf{and} }

\usepackage[noadjust]{cite}

\usepackage{amsmath,amssymb,amsfonts}
\usepackage{mathabx}
\usepackage{mathtools}
\usepackage{graphicx}
\usepackage{textcomp}
\usepackage{url}
\usepackage{caption, subcaption}
\usepackage{hyperref}
\usepackage{svg}
\usepackage[normalem]{ulem}
\useunder{\uline}{\ul}{}
\usepackage{float} 
\usepackage{listings}[language=C++]
\usepackage{multirow}
\usepackage{enumitem}

\usepackage{amsthm}
\usepackage[utf8]{inputenc}
\theoremstyle{plain}

\newtheorem{lemma}{Lemma}
\theoremstyle{definition}
\newtheorem{definition}{Definition}
\newtheorem{example}{Example}

\newcommand{\red}[1]{{{\color{red} #1}}}

\usepackage{listings}
\usepackage{xcolor}

\definecolor{codegreen}{rgb}{0,0.6,0}
\definecolor{codegray}{rgb}{0.5,0.5,0.5}
\definecolor{codepurple}{rgb}{0.58,0,0.82}
\definecolor{backcolour}{rgb}{0.95,0.95,0.92}

\lstdefinestyle{mystyle}{
    backgroundcolor=\color{backcolour},   
    commentstyle=\color{codegreen},
    keywordstyle=\color{magenta},
    numberstyle=\tiny\color{codegray},
    stringstyle=\color{codepurple},
    basicstyle=\ttfamily\footnotesize,
    breakatwhitespace=false,         
    breaklines=true,                 
    captionpos=b,                    
    keepspaces=true,                 
    numbers=left,                     
    numbersep=5pt,                  
    showspaces=false,                
    showstringspaces=false,
    showtabs=false,                  
    tabsize=2
}

\usepackage{pifont}

\title{GCAPS: GPU Context-Aware Preemptive Priority-based Scheduling for Real-Time Tasks}
\titlerunning{GCAPS: GPU Preemptive Real-Time Task Scheduling}

\author{Yidi Wang}{University of California, Riverside, USA}{ywang665@ucr.edu}{}{}
\author{Cong Liu}{University of California, Riverside, USA}{congl@ucr.edu}{}{}
\author{Daniel Wong}{University of California, Riverside, USA}{danwong@ucr.edu}{}{}
\author{Hyoseung Kim}{University of California, Riverside, USA}{hyoseung@ucr.edu}{}{}

\authorrunning{Y. Wang, C. Liu, D. Wong, and H. Kim}

\Copyright{Yidi Wang, Cong Liu, Daniel Wong, and Hyoseung Kim}



\acknowledgements{This work is supported by the National Science Foundation (NSF) grants 1943265, 1955650, and 2312395.}

\nolinenumbers

\EventEditors{Rodolfo Pellizzoni}
\EventNoEds{1}
\EventLongTitle{36th Euromicro Conference on Real-Time Systems (ECRTS 2024)}
\EventShortTitle{ECRTS 2024}
\EventAcronym{ECRTS}
\EventYear{2024}
\EventDate{July 9--12, 2024}
\EventLocation{Lille, France}
\EventLogo{}
\SeriesVolume{298}
\ArticleNo{11}

\begin{document}

\maketitle

\begin{abstract}
Scheduling real-time tasks that utilize GPUs with analyzable guarantees poses a significant challenge due to the intricate interaction between CPU and GPU resources, as well as the complex GPU hardware and software stack. While much research has been conducted in the real-time research community, several limitations persist, including the absence or limited availability of GPU-level preemption, extended blocking times, and/or the need for extensive modifications to program code. In this paper, we propose GCAPS, a \underline{G}PU \underline{C}ontext-\underline{A}ware \underline{P}reemptive \underline{S}cheduling approach for real-time GPU tasks. Our approach exerts control over GPU context scheduling at the device driver level and enables preemption of GPU execution based on task priorities by simply adding one-line macros to GPU segment boundaries. In addition, we provide a comprehensive response time analysis of GPU-using tasks for both our proposed approach as well as the default Nvidia GPU driver scheduling that follows a work-conserving round-robin policy. Through empirical evaluations and case studies, we demonstrate the effectiveness of the proposed approaches in improving taskset schedulability and response time. The results highlight significant improvements over prior work as well as the default scheduling approach, with up to 40\% higher schedulability, while also achieving predictable worst-case behavior on Nvidia Jetson embedded platforms.
\end{abstract}

\section{Introduction}

Real-time cyber-physical systems with GPU workloads have become increasingly prevalent in various domains including self-driving cars, autonomous robots, and edge computing nodes. This trend has been accelerated in recent years by the demand for learning-enabled components as most of their implementations heavily rely on the GPU stack. The scheduling problem of GPU-using tasks in these systems is therefore crucial to ensure timely execution and to meet stringent timing requirements. One of the key challenges here is effectively supporting prioritization and preemption, allowing higher-priority tasks to interrupt and supersede lower-priority GPU tasks whenever needed. This is particularly important in scenarios where critical high-priority tasks with stringent deadlines need to access GPU resources, while low-priority and best-effort tasks can tolerate such preemption to accommodate their execution. 

As of yet, the default scheduling policy of commercial GPU devices provides little control over the prioritization and preemption of GPU access segments of tasks, causing unpredictable task response time and instability in real-time systems. 
The real-time research community has recognized this issue since the early era of GPU computing and has proposed several solutions. In particular, the use of real-time synchronization protocols, such as MPCP~\cite{rajkumar1990real,patel2018analytical} and FMLP+~\cite{brandenburg2014fmlp+}, has been recognized as a promising way to manage GPU tasks in real-time systems with strong analyzable guarantees on the worst-case task response time. However, these approaches can suffer from long blocking time and priority inversion by lower-priority tasks since GPU segments are handled non-preemptively. There have been attempts to support priority-based GPU scheduling with preemption capabilities~\cite{Kato2011_RGEM, Basaran2012, Zhou2015}, but they require significant modifications to GPU access code, lack analytical support, and more importantly, may not work properly if the system has processes with unmodified GPU code or graphics applications due to the time-shared GPU context switching behavior of the device driver~\cite{capodieci2018deadline,Bakita2023}. 

In this paper, we address the aforementioned challenges and limitations by proposing GCAPS, \underline{G}PU \underline{C}ontext-\underline{A}ware \underline{P}reemptive \underline{S}cheduling, for real-time GPU task execution in multi-core systems with analyzable guarantees. 
Our work focuses on Nvidia GPUs, especially those on Tegra System-on-Chips (SoCs) used in embedded platforms like Jetson Xavier and Orin running L4T (Linux for Tegra). 
The proposed approach works at the device driver level, and unlike existing techniques, they can protect the execution of real-time GPU processes from interference from best-effort non-real-time CUDA processes and graphics processes in the system. Specifically, compared to the existing approaches to enable GPU preemption, our approach requires minimum modifications to the user-level GPU access code, i.e., adding just one macro at the boundaries of GPU segments, but provides more fine-grained and efficient control of the GPU. This is particularly appealing to recent machine learning and computer vision applications as they are built on top of massive libraries that involve hundreds of different kernels.
Thanks to the strictly preemptive and priority-driven GPU scheduling behavior, the proposed approach is analyzable and allows us to derive response-time tests for schedulability analysis.

In summary, the paper makes the following contributions:
\begin{itemize}
    \item We propose a novel GPU context-aware preemptive priority-driven GPU scheduling approach for a multi-core system equipped with an Nvidia GPU. This approach not only enables GPU segments to be executed according to their task priority (especially important  when task priority is assigned based on criticality), but also provides a way to assign different priorities to GPU segments, which yields a significant benefit in schedulability. 
    \item We present a comprehensive analysis on the worst-case task response time under our proposed approach. In particular, our analysis considers both self-suspension and busy-waiting modes during GPU kernel execution, as well as the overhead caused by GPU context switching, which has been neglected in the literature. We also analyze the response time of a GPU-using task under the vanilla Nvidia Tegra GPU driver that follows a work-conserving round-robin policy.
    \item Our work is implemented on the latest Nvidia Tegra driver and will be open-sourced.
    \footnote{Available at \url{https://github.com/rtenlab/gcaps-super-repo}}. 
    Experimental results show that our approaches bring substantial benefits in taskset schedulability compared to previous synchronization-based approaches. A case study on Jetson Xavier and Orin platforms demonstrates the effectiveness of our work over the default GPU driver and the applicability to various generations of GPU architectures.
\end{itemize}

\section{Background on Tegra GPU Scheduling}\label{sec:background}
Computational GPU workloads for Nvidia GPUs are often programmed using the CUDA library. These workloads are represented in \textit{kernels} and user-level processes can launch kernels to the GPU at runtime. CUDA provides processes with \textit{streams} to enable concurrent execution of kernels with a limited number of stream priority levels, e.g., only 2 in the Pascal architecture~\cite{xiang2019pipelined}. 

Since streams are bound to a user-level process that created them, the effect of stream scheduling and stream priority assignment is exerted only within each process boundary. 
The CUDA library is not a must for processes to access the GPU hardware. There are other low-level libraries for general-purpose GPU computing and graphics applications such as OpenCL and Vulkan. Programs built using different libraries co-exist in the system and they send GPU commands to the device driver.

\noindent\textbf{Time-shared scheduling. }
At the device driver level, each process is associated with a \textit{GPU context}, which represents a virtual address space and other runtime states on the GPU side. Any process accessing the GPU has a separate GPU context, regardless of whether it uses the CUDA library or not in the user space, and GPU contexts from different processes are time-sliced to share the GPU hardware.

\begin{figure}[]
\vspace{-10pt}
    \centering
    \includegraphics[width=0.7\linewidth]{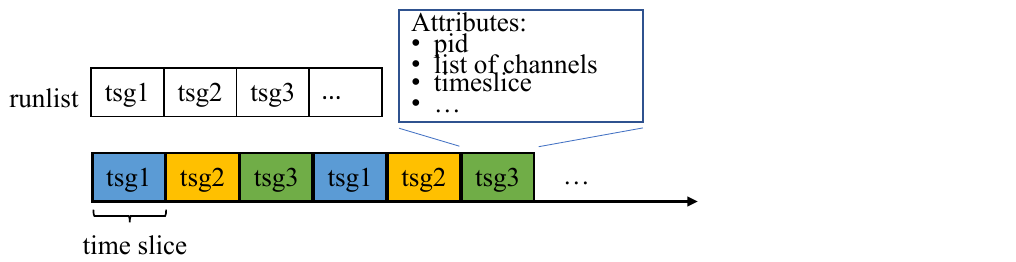}
    \caption{Runlist and time-sliced GPU scheduling}
    \label{fig:runlist_and_tsgs}
\end{figure}

To ensure fairness and prevent resource contention, the Tegra GPU driver uses a scheduling policy that assigns entries in the ``runlist''.
\footnote{In fact, there are multiple runlists but we refer to them as singular for simplicity. Each runlist corresponds to a specific hardware engine, such as copy engine or graphic engine. By default, every process maintains a single TSG entry on each runlist for storing the commands to be executed by that respective engine. This configuration does not impact the structure of our proposed design.}
The entries of the runlist represent the allocation of time slices to TSGs (Time-Sliced Groups~\cite{bakita2024demystifying}) that are directly associated with processes. Each TSG has multiple ``channels'', each of which contains a stream of GPU commands received from its process. 
The runlist is populated with entries of TSGs, as depicted in Fig.~\ref{fig:runlist_and_tsgs}. Each TSG entry maintains state attributes like the process ID, a list of channels, and the allocated time slice.

\noindent\textbf{Runlist construction. }
The runlist is constructed by processes submitting commands to their respective TSG channels. Specifically, as commands are submitted, TSG entries are added to the runlist, which is managed under a mutex lock to prevent race condition. 
The device driver can assign priority to TSGs, and TSGs with higher driver-level priority are allocated larger time slices and more entries on the runlist. 
Following its construction, the runlist is scheduled by the GPU in a \textit{round-robin} fashion, where each TSG entry's commands are executed for up to its time slice before moving to the next entry. This procedure continues until all commands of all active TSGs on the runlist have been executed.

As of this writing, there is no interface provided to the user to configure the length of time slices or the TSG priority setting. We have observed that the latest Tegra driver uses the same length of time slices for all TSGs, implying that the default scheduling policy of the driver focuses on ensuring fairness across different processes accessing the GPU.

\noindent\textbf{GPU context switching. }
Moving from one TSG to another on the runlist causes GPU context switching. The major contributors to GPU context switching overhead are register file saving and cache flushing, the former taking much longer than CPUs due to the GPU's large register files~\cite{Tanasic2014}.\footnote{Since the Pascal architecture, Nvidia GPUs use demand paging for GPU memory management. Thus, the GPU memory of a context is not swapped out during GPU context switching, regardless of whether the GPU is integrated or discrete.} In addition, extra delay may occur because Nvidia GPUs support preemption at the pixel level for graphics tasks and the thread-block level for compute tasks~\cite{nvidia_preemption}. In the case of data copy operations, data is divided into smaller chunks and preemption occurs at the boundary of each chunk~\cite{capodieci2018deadline}. Such delay is however very small compared to the length of GPU kernels, and for compute tasks, it can be separately measured or estimated by the maximum length of a single thread block among all kernels. Hence, we define the following term: 
\begin{definition}[GPU context switch overhead]	\label{def:ctxsw_overhead_theta}
	The GPU context switch overhead, $\theta$, is the time required to switch from the GPU context of one process to that of another process, including all the aforementioned delay factors. 
\end{definition}

Prior work~\cite{capodieci2018deadline} reports that GPU context switching can take from 50 to 750~$\mu$s, which can be estimated by considering the GPU cache size and memory access latency. Our measurements in Sec.~\ref{sec:system_eval} show similar results, and our analysis in Sec.~\ref{sec:analysis} accounts for this overhead with $\theta$. 

\smallskip
In summary, the Tegra GPU driver employs a \textit{time-sliced round-robin} scheduling approach. This approach, however, does not respect the OS-level scheduling priority of processes, which is the main control knob to tune real-time performance in practice. This led to diminished responsiveness in high-priority real-time tasks whenever the system accommodated new low-priority or best-effort tasks. In addition, it is not easy for the user to observe such driver-level behavior because GPU profiling tools, such as Nvidia Nsight Systems, do not report GPU context switching events and each kernel execution time appears to be inflated with no time slice information. These issues contribute to difficulties in understanding and predicting the runtime behavior of GPU-enabled real-time systems.

\begin{table}[]
\small
\tiny\centering
\begin{tabular}{|l|l|l|l|l|l|}
\hline
&                                                                          & \begin{tabular}[c]{@{}l@{}}No \\ blocking\end{tabular} & \begin{tabular}[c]{@{}l@{}}Task priority\\ respected\end{tabular} & \begin{tabular}[c]{@{}l@{}}Analyzable\\ response time\end{tabular} & \begin{tabular}[c]{@{}l@{}}Inter-GPU \\ context\end{tabular} \\ \hline
\multirow{4}{*}{\begin{tabular}[c]{@{}l@{}}Prior\\ Work\end{tabular}} & \begin{tabular}[c]{@{}l@{}}Unmanaged GPU\\ (default driver)\end{tabular} & \ding{53}   & \ding{53}                                                         & \ding{51}                                                          & \ding{51}                                                    \\ \cline{2-6} 
& \begin{tabular}[c]{@{}l@{}}Sync.-based approaches \\ \cite{Elliott2012,Elliott_RTS13,Elliott_RTSS13,patel2018analytical}      \end{tabular}                                            & \ding{53}   & \ding{51}                                                         & \ding{51}                                                          & \ding{51}                                                    \\ \cline{2-6} 
& \begin{tabular}[c]{@{}l@{}}GPU partitioning\\ \cite{Bakita2023,Saha2019,wang2021balancing,Jain2019,wang2022towards,zou2023rtgpu,wu2015enabling}\end{tabular}    & \ding{51}   & \ding{53}                                                         & \ding{51}                                                          & \ding{53}                                                    \\ \cline{2-6} 
& \begin{tabular}[c]{@{}l@{}}Preemptive GPU \\ \cite{Kato2011_RGEM,Basaran2012,Zhou2015,capodieci2018deadline,Han2022_reef}     \end{tabular}                                          & \ding{51}   & \ding{51}                                                         & Unknown                                                                  & \ding{51}                                                    \\ \hline
Ours    & GCAPS                                                     & \ding{51}   & \ding{51}                                                         & \ding{51}                                                          & \ding{51}                                                    \\ \hline
\end{tabular}
\caption{Comparison of different GPU scheduling approaches}
\label{tab:comparison_with_prior_work}
\end{table}

\section{Related Work}\label{sec:related_work}

Table~\ref{tab:comparison_with_prior_work} gives a summary of comparison between representative GPU scheduling approaches. Below we discuss prior work in various categories.

\noindent\textbf{Synchronization-based GPU access control.} 
Real-time synchronization protocols have played an important role in managing access to GPUs~\cite{Elliott2012,Elliott_RTS13,Elliott_RTSS13,patel2018analytical}. With this approach, GPUs are modeled as mutually-exclusive shared resources and tasks are made to acquire locks to enter code segments accessing the GPUs, i.e., critical sections. MPCP~\cite{rajkumar1990real} and FMLP+~\cite{brandenburg2014fmlp+} are prime examples for multi-core systems with GPUs and the use of such protocols enables analytically provable worst-case task response time bounds.
However, as we discussed in Sec.~\ref{sec:background}, those works overlooked the inherent TSG context switching overhead. Another drawback is that the synchronization-based approach may suffer from blocking time from lower-priority tasks holding a lock and priority inversion caused by the priority boosting mechanism employed in these protocols~\cite{HKim2017}. This becomes particularly problematic when tasks busy-wait on long kernel execution, as discussed in~\cite{patel2018analytical}. 

\noindent\textbf{Preemptive GPU scheduling.} Several previous studies~\cite{Kato2011_RGEM, Basaran2012, Zhou2015} have proposed software-based mechanisms to enable preemptive scheduling of real-time GPU tasks. These approaches introduce the concept of decomposing long-running GPU kernels into smaller blocks, allowing preemption to occur at the boundaries of these blocks. By enabling preemptive scheduling, the waiting time of high-priority tasks can be significantly reduced, improving responsiveness and offering a better chance to meet timing requirements. However, the cost of utilizing these mechanisms is not trivial as they necessitate a significant rewriting of user programs~\cite{Basaran2012} or an implementation of a custom CUDA library with device driver modifications~\cite{Basaran2012,Zhou2015}.
Capodieci et al.~\cite{capodieci2018deadline} proposed a hypervisor-based technique to support preemptive Earliest Deadline First (EDF) GPU scheduling of virtual machines (VMs) in a virtualized environment. This approach achieves GPU performance isolation among VMs and shares some similarities with our work, in terms of controlling GPU context switching at the device driver level. However, it lacks consideration of the end-to-end response time of tasks involving CPU and GPU interactions, which is a specific focus of our work.
Recently, Han et al.~\cite{Han2022_reef} proposed REEF, which enables microsecond-scale, reset-based preemption for concurrent DNN inferences on GPUs. This approach proactively kills and restarts best-effort kernels leveraging the idempotent nature of most DNN inference, but it is not applicable to a wide range of applications.

\noindent\textbf{GPU partitioning.} As a GPU is composed of multiple compute units, e.g., Streaming Multiprocessors (SMs) on Nvidia GPUs, there have been attempts to spatially partition the GPU and make them accessible by multiple real-time tasks in parallel~\cite{Saha2019,wang2021balancing,Jain2019,wang2022towards,zou2023rtgpu}. They use SM-centric kernel transformation~\cite{wu2015enabling} to run kernels on their designated SMs/partitions. As this involves extensive program modifications and may suffer from misbehaving tasks, Bakita and Anderson~\cite{Bakita2023} recently proposed a user-space library that minimizes program changes and offers much better usability and portability. With GPU partitioning, task performance is greatly affected by partitioning results, e.g., a high-priority task may suffer performance degradation due to the small number of SMs assigned to it or experience blocking if its SMs are shared with other tasks. In addition, 
all these approaches work within a single GPU context, i.e., one process; hence, multiple processes with separate contexts will still time-share the GPU, as discussed in Sec.~\ref{sec:background}. Note that our work does not compete with GPU partitioning techniques. They can be used within each process and our work enables predictable scheduling of GPU processes.



\section{System Model}

We consider a multi-core system with a GPU, which is common in today's embedded hardware platforms like Nvidia Jetson. The CPU has $\omega$ identical cores and the GPU is yet another processing resource used by compute-intensive tasks. The GPU consists of internal resources including Execution Engines (EEs) and Copy Engines (CEs). 
The EE and CE operations of a single process can be done asynchronously at runtime, and during pure GPU execution, the process can either busy-wait or self-suspend on the CPU. However, different processes cannot use the GPU at the same time because of the time-sharing scheduling of GPU contexts at the GPU device driver, as discussed before. 

\smallskip\noindent\textbf{Task Model. }
We consider a taskset $\Gamma$ consisting of $n$ sporadic tasks (processes) with fixed priority and constrained deadlines.\footnote{We assume tasks are processes and use them interchangeably in this paper.} Out of these, $n^g$ tasks require GPU operations.
Each task is assumed to be preallocated to one CPU core with no runtime migration, i.e., {\em partitioned multiprocessor scheduling}.
The execution of a task is an alternating sequence of CPU segments and GPU segments. CPU segments run entirely on the CPU and GPU segments involve GPU operations such as memory copy and kernel execution. 
A task $\tau_i$ can be characterized as follows:
\begin{equation*}
\small
    \tau_i := (C_i, G_i, T_i, D_i, \eta^c_i, \eta^g_i, \pi_i)
\end{equation*}

\begin{itemize}
    \item $C_i$: the cumulative sum of the worst-case execution time (WCET) of all CPU segments of task $\tau_i$.
    \item $G_i$: the cumulative WCET of GPU segments (including memory copies and kernels) of $\tau_i$.
    \item $T_i$: the minimum inter-arrival time of each job of $\tau_i$.
    \item $D_i$: the relative deadline of each job of $\tau_i$, assumed to be smaller than or equal to the period, i.e., $D_i \le T_i$.
    \item $\eta^c_i$: the number of CPU segments in each job of task $\tau_i$.
    \item $\eta^g_i$: the number of GPU segments in each job of task $\tau_i$; if $\tau_i$ does not use the GPU, $\eta^g_i=0$.
    \item $\pi_i$: the priority of task $\tau_i$.\footnote{Our work allows the GPU segments of $\tau_i$ to run with a separate priority from its OS-level process priority to improve schedulability (explained later in Sec.~\ref{sec:separate_gpu_prio}). We will use $\pi^c_i$ for CPU segment priority and $\pi^g_i$ for GPU segment priority. If not specified, $\pi_i=\pi^c_i=\pi^g_i$.} 
\end{itemize}

\begin{figure}[h]
\vspace{-10pt}
\centering
    \includegraphics[width=0.8\linewidth]{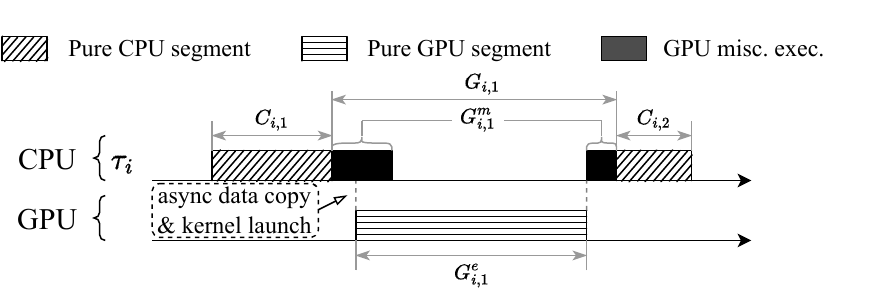}
    \vspace{-0.5\baselineskip}\caption{Task model}
\label{fig:task_model_orig}
\end{figure}
\vspace{-10pt}
Fig.~\ref{fig:task_model_orig} depicts these parameters, and by default, in each task, all the segments have the same priority. 
Each GPU segment $G_{i,j}$\footnote{Although Tegra GPU supports zero-copy memory transfer, we still include the memory copy in the diagrams to illustrate the complete process. It is important to note that the choice of using zero-copy or not does not impact the analysis or the experimental results of this work.} can be characterized as follows:
\begin{equation*}
\small
    G_{i,j} := (G^m_{i,j}, G^e_{i,j}) 
\end{equation*}

\begin{itemize}
    \item $G^m_{i,j}$: the cumulative WCET of miscellaneous CPU operations in the $j$-th GPU segment of task $\tau_i$, $G_{i,j}$.
    \item $G^e_{i,j}$: the WCET of GPU workload in $G_{i,j}$ that requires {\em no} CPU intervention such as data copy and kernel execution; and we call it a {\em pure GPU segment}.
\end{itemize}

$G^m_{i,j}$ is the time for launching a CUDA kernel, overhead for communicating with the GPU driver, and miscellaneous CPU operations for issuing other GPU commands. 
$G^e_{i,j}$ is the time for GPU data copy and kernel execution, during which task $\tau_i$ can either busy-wait or self-suspend on the CPU.
Note that $G_{i,j} \leq G^m_{i,j} + G^e_{i,j}$ because the worst-case of $G^m$ and $G^e$ are not necessarily happening on the same control path and they may execute in parallel in asynchronous mode~\cite{patel2018analytical}.


\section{GCAPS: Priority-based Preemptive GPU Context Scheduling}\label{sec:priority_based_preemptive_gpu_scheduling}

This section presents our priority-based preemption GPU context-aware scheduling approach. It involves a set of user-level runtime macros that notify the GPU driver to update the runlist, and it provides fine-grained control over GPU segments. 

\subsection{GCAPS Algorithm}\label{sec:gcaps_algorithm}
We first introduce the high-level scheduling procedures of GCAPS.
To implement the approach, we add two macros that allow user programs to indicate the beginning and completion of a GPU segment. When the macro is called, it generates an IOCTL command and sends it to the GPU driver through a file descriptor, and requests the driver to update the runlist accordingly. While we chose IOCTL as a way to interact with the driver, other methods could be used as well, such as system calls and \texttt{procfs}/\texttt{sysfs} interfaces. 

The macros introduced are \texttt{gcapsGpuSegBegin()} and \texttt{gcapsGpuSegEnd()}, which are wrappers to our IOCTL syscalls. A sample user program is listed in Listing.~\ref{lst:ioctl_sample}. The code between them is a GPU segment. 
With the help of these two macros, we can let our driver-level approach know the boundaries of GPU segments and make GPU scheduling decisions at the right time.

\begin{lstlisting}[caption={Example usage of GCAPS macros}, label={lst:ioctl_sample}, captionpos=t, float, abovecaptionskip=-\medskipamount, language=C++, aboveskip=-.5\baselineskip, belowskip=0\baselineskip]
int task_function() {
    ...
    gcapsGpuSegBegin(fd, getpid()); /* GCAPS: GPU segment begin */
    cudaMemcpyAsync(d_in, h_in, mem_size_in, cudaMemcpyHostToDevice, stream));
    MyKernel<<<grid, threads, 0, stream>>>(d_in, d_out);
    cudaMemcpyAsync(h_out, d_out, mem_size_in, cudaMemcpyHostToDevice, stream));
    gcapsGpuSegEnd(fd, getpid()); /* GCAPS: GPU segment finish */
    ...
}
\end{lstlisting}

\begin{algorithm}[t]
\footnotesize
\begin{algorithmic}[1]  
\State $task\_pending = \emptyset$
\State $task\_running = \emptyset$
\State \Comment{Note that a task exclusively exists in one of these two lists}
\Procedure{TSG\_Scheduler}{$\tau_i, add$}        
    \If{$add$} \Comment{$\tau_i$ requests to be added}
        \If{$\tau_i$ is not a real-time task} \label{alg2:tau_i_not_rt}
            \If{no real-time task is in $task\_running$}
                \State Add $\tau_i$ to $task\_running$
            \Else
                \State Add $\tau_i$ to $task\_pending$ \label{alg2:tau_i_not_rt_done}
            \EndIf
        \Else \Comment{$\tau_i$ is a real-time task} \label{alg2:tau_i_rt}
            \State $\tau_h \leftarrow$ the highest-priority task in $task\_running$ \label{alg2:tau_h_assigned_highest_prio}
            \If{$\tau_i.rt\_priority > \tau_h.rt\_priority$} \label{alg2:comp_prio}
                \State Add $\tau_i$ to $task\_running$
                \State Move $\tau_h$ to $task\_pending$
            \Else
                \State Add $\tau_i$ to $task\_pending$ \label{alg2:tau_i_rt_done}
            \EndIf
        \EndIf
    \Else \Comment{$\tau_i$ requests to be removed} \label{alg2:tau_i_to_be_removed}
        \State $\tau_k \leftarrow$ the highest-priority RT task in $task\_pending$ \label{alg2:tau_k_assigned_highest_prio}
        \If{$\tau_k$ exists}
            \State Move $\tau_k$ to $task\_running$
            \State Remove $\tau_i$ from $task\_running$
        \Else \Comment{no pending real-time task}
            \State $task\_running \gets task\_pending$
            \State $task\_pending \gets \emptyset$ \label{alg2:tau_i_to_be_removed_done}
        \EndIf
    \EndIf
    \State Add all TSGs of tasks in $task\_running$ to the runlist
\EndProcedure
\end{algorithmic}  
\caption{Priority-based TSG scheduling}
\label{alg:ioctl_approach}  
\end{algorithm}  

An IOCTL command issued by the macro triggers our TSG scheduler shown in Alg.~\ref{alg:ioctl_approach}. 
To keep track of which tasks are in the runlist and which tasks are pending, two bitfield lists, $task\_running$ and $task\_pending$, are maintained in the GPU driver.
When a caller task $\tau_i$ notifies that it begins its GPU segment through \texttt{gcapsGpuSegBegin()}, the scheduler first checks whether $\tau_i$ is a real-time task by checking whether the \texttt{rt\_priority} field of the task's \texttt{task\_struct} is set. If it is not, the scheduler checks whether there is any real-time task that is currently running and decides whether to add $\tau_i$ to the runlist or add it to the pending list (line~\ref{alg2:tau_i_not_rt} to~\ref{alg2:tau_i_not_rt_done}). If $\tau_i$ is a real-time task, the scheduler checks the priority of $\tau_i$ relative to the currently-running highest-priority task $\tau_h$ ($\tau_h \in task\_running \land \tau_h \neq \tau_i$). If the priority of $\tau_i$ is higher than $\tau_h$, the scheduler preempts the GPU execution of $\tau_h$ and moves it to the pending list, and $\tau_i$ is added to the runlist. Otherwise, $\tau_i$ is added to the pending list (line~\ref{alg2:tau_i_rt} to~\ref{alg2:tau_i_rt_done}).
If $\tau_i$ notifies the driver about the completion through \texttt{gcapsGpuSegEnd()}, the scheduler first finds the highest-priority task $\tau_k$ in the pending list. If $\tau_k$ exists, it is added to the runlists. Otherwise, if there are only best-effort tasks, the scheduler adds all of them to the runlist to resume their progress in a time-shared manner (line~\ref{alg2:tau_i_to_be_removed} to~\ref{alg2:tau_i_to_be_removed_done}). At the end of the scheduler, it directly updates the driver's runlist based on $task\_running$ such that the TSGs of tasks are dispatched and GPU context switching takes place immediately. To implement the scheduler, we only added about 300 lines of code in the driver, and 10 lines of code in each userspace macro.

\begin{figure}[t]
\centering
    \begin{subfigure}[b]{0.8\linewidth}
    \vspace{-10pt}
        \centering
        \includegraphics[width=\linewidth]{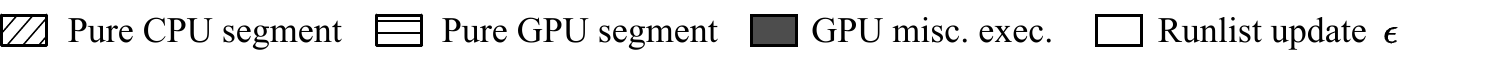}
    \end{subfigure}\\

\begin{minipage}{1\linewidth}
    \hspace{-20pt}
    \begin{subfigure}[b]{0.52\linewidth}
    \centering
        \includegraphics[width=\linewidth]{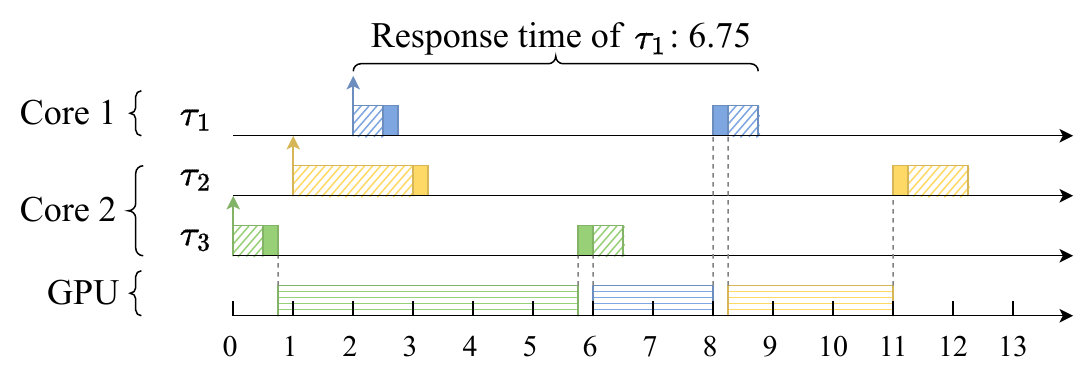}
        \vspace{-1\baselineskip}\caption{Synchronization-based approach}
        \label{fig:sync_based_approach}
    \end{subfigure}
    \begin{subfigure}[b]{0.52\linewidth}
    \centering
        \includegraphics[width=\linewidth]{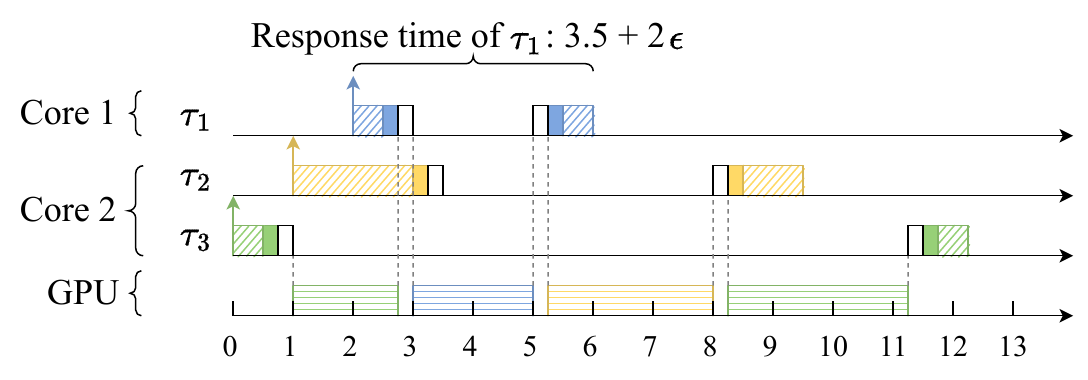}
        \vspace{-1\baselineskip}\caption{Proposed approach}
        \label{fig:ioctl_approach}
    \end{subfigure}
\end{minipage}
\caption{Example schedule of three tasks under different approaches (priority $\tau_1 > \tau_2 > \tau_3$)}
\label{fig:motivation_example}
\end{figure}

\begin{example}[Motivational example]
Figs.~\ref{fig:motivation_example} compares task schedules under the conventional synchronization-based approach and our proposed approach. $\tau_1$ is running on Core 1, while $\tau_2$ and $\tau_3$ are running on Core 2. 
The synchronization-based approach shown in Fig.~\ref{fig:sync_based_approach} treats the entire execution of a GPU segment as a critical section. Tasks are serviced in order based on their task priorities. This approach ensures that each task completes its GPU segments in a deterministic and predictable manner.
However, $\tau_1$ is delayed by the GPU segments of all of its lower-priority tasks and gets a response time of 6.75. 

On the other hand, our approach avoids this delay by allowing preemption during GPU segment execution. Fig.~\ref{fig:ioctl_approach} shows an example schedule under the proposed approach, and the overhead introduced by our approach is denoted as $\epsilon$, defined as follows.
\end{example}

\begin{definition}[Runlist update delay]\label{def:runlist_update_delay}
The runlist update delay, $\epsilon$, is defined as the sum of the time it takes to complete our TSG scheduler (represented by $\alpha$, including the cost for IOCTL system call, TSG scheduling algorithm, and runlist update) and the resulting GPU context switching overhead ($\theta$). Hence, $ \epsilon = \alpha+\theta$. 
\end{definition}

Unlike the synchronization-based approach, at $t=3-\epsilon$, $\tau_1$'s GPU segment issues an IOCTL syscall to notify the driver that its GPU segment is ready to run. It causes preemption of $\tau_3$'s GPU segment by removing its associated TSGs from the runlist, and the GPU is solely occupied by the highest-priority task in the system, $\tau_1$, until it completes. The response time of $\tau_1$ is $3.5+2\epsilon$, much smaller than that of the synchronization-based approach.
This strategy is followed in the remaining schedule.

\begin{figure}[t]
\vspace{-10pt}
\centering
\begin{minipage}{0.52\linewidth}
\vspace{-10pt}
    \begin{subfigure}{\linewidth}
        \captionsetup{justification=centering}
        \includegraphics[width=\linewidth]{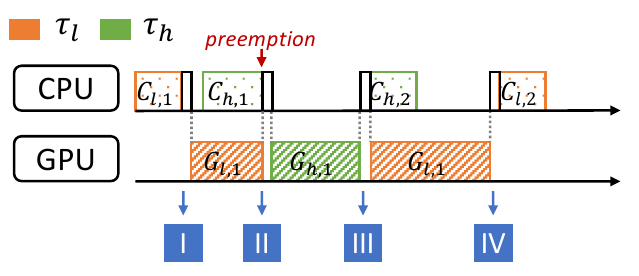}
        \vspace{-1.5\baselineskip}\caption{Preemption trigger points ($G^m$ is omitted)}
        \label{fig:flow_timeline}
    \end{subfigure}
    \begin{subfigure}{\linewidth}
        \captionsetup{justification=centering}
        \includegraphics[width=\linewidth]{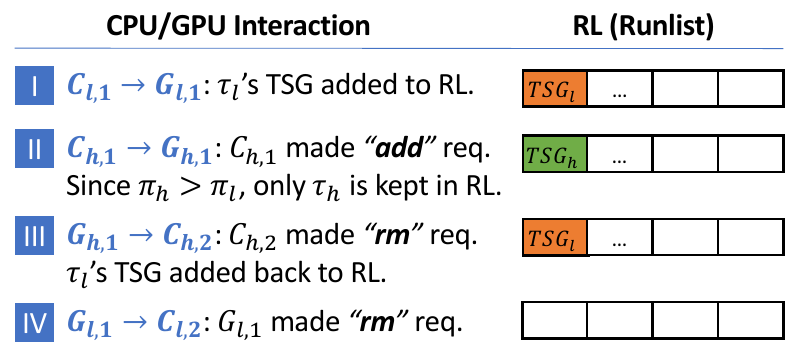}
        \vspace{-1\baselineskip}\caption{GPU/CPU Interaction of Fig.~\ref{fig:flow_timeline}}
        \label{fig:flow_cpu-gpu-interaction}
    \end{subfigure}
\end{minipage}%
\hfill
\begin{minipage}{0.47\linewidth}
    \begin{subfigure}[b]{\linewidth}
        \captionsetup{justification=centering}
        \centering
        \includegraphics[width=\linewidth]{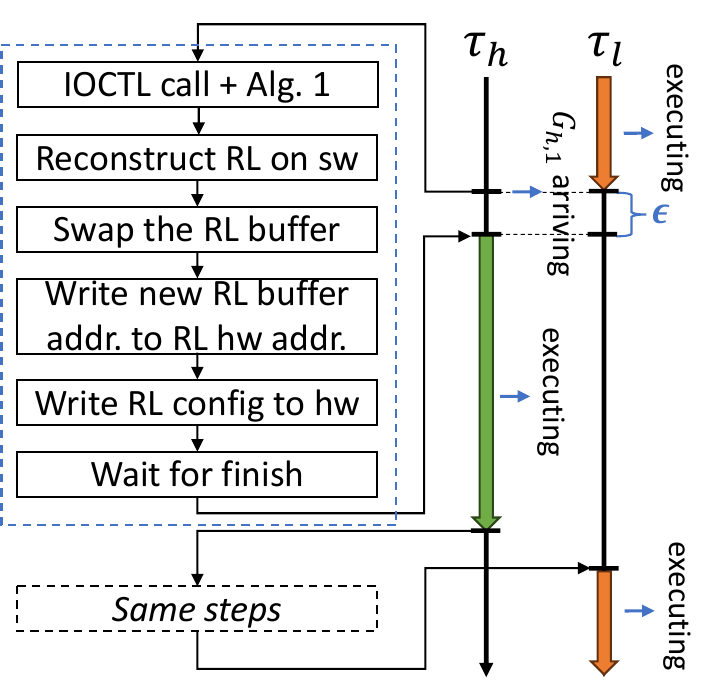}
        \vspace{-1\baselineskip}\caption{Context Switching Details}
        \label{fig:flow_ctxsw}
    \end{subfigure}
\end{minipage}%
\caption{Preemption by GPU segments and GPU context switching}
\end{figure}

\subsection{GPU Context Switching Details}\label{sec:gcaps_ctxsw_details}
This section details the context switching of GPU segments when preemption is triggered. Fig.~\ref{fig:flow_timeline} shows a schedule of two real-time tasks, where preemption points are marked in red and the CPU and GPU interactions are marked in blue. Fig.~\ref{fig:flow_cpu-gpu-interaction} illustrates the runlist status at each CPU/GPU interaction. Note that, even if the active TSGs of a task are removed from the runlist, they are kept in the scheduler data structure of the GPU driver and won't be lost; hence, we can add those TSGs back to the runlist, e.g., $G_{h,1} \rightarrow C_{h,2}$ in Fig.~\ref{fig:flow_cpu-gpu-interaction}, to resume their execution.

Fig.~\ref{fig:flow_ctxsw} illustrates the detailed steps of GPU context switching between $\tau_l$ and $\tau_h$. The procedures outlined within the dashed blue block represent the entire cost of preemption ($\epsilon$).
After the IOCTL system call is invoked, Alg.~\ref{alg:ioctl_approach} is executed in the driver and it identifies the ids of all the TSGs for the runlist. 
To manage concurrent system calls efficiently in our proposed approach, we replace the default mutex lock in the driver with a real-time mutex, \texttt{rt\_mutex}~\cite{rt_mutex}, to reduce the blocking time as well as prevent priority inversion. 

Once the new runlist is constructed, the scheduler swaps it with the one currently held by hardware. This is similar to the well-known double buffering technique. 
The scheduler allocates runlist buffers in DMA memory during the driver initialization phase.
The size of each buffer corresponds to the product of the entry size and the number of entries, which depend on hardware capabilities. For instance, on Nvidia Jetson Xavier NX, each entry is 16 bytes with 65535 entries in total.
Then the runlist buffer to use is submitted to the hardware. It involves writing the new runlist buffer address to the runlist's hardware address and writing the runlist configurations to the hardware registers. During the submission, the driver polls the hardware and waits until it finishes.
After these operations, the new runlist only contains the TSGs of $\tau_h$ in Fig.~\ref{fig:flow_ctxsw}, and $\tau_h$ can run in isolation without interference from $\tau_l$.

\subsection{Separate Priority for GPU Segments}
\label{sec:separate_gpu_prio}

\begin{table}[]
\vspace{-10pt}
\small
\centering
\begin{tabular}{l|l|l|l|l}
\hline
Task     & CPU & $T_i=D_i$   & CPU Segments                          & GPU Segments                                                                                         \\ \hline
$\tau_1$ & 1   & 80 & $C_{1,1}=2, C_{1,2}=4, C_{1,3}=3$ & $G^m_{1,1}=2, G^e_{1,1}=4$, $G^m_{1,2}=2, G^e_{1,2}=2$ \\
$\tau_2$ & 1   & 150 & $C_{2,1}=40$                      & -                                                                                                \\
$\tau_3$ & 2   & 190 & $C_{3,1}=4, C_{3,2}=30$          & $G^m_{3,1}=5, G^e_{3,1}=80$                                                                     \\
$\tau_4$ & 1   & 200 & $C_{4,1}=16, C_{4,2}=2$           & $G^m_{4,1}=2, G^e_{4,1}=10$                                                                       \\ \hline
\end{tabular}
\caption{Taskset used in Fig.~\ref{fig:gpu_prio_example}}
\label{tab:taskset_separate_priority_example}
\end{table}

\begin{figure}[t]
\vspace{-10pt}
\centering
    \begin{subfigure}[b]{0.8\linewidth}
        \centering
        \includegraphics[width=\linewidth]{figs/examples/universal_legend.pdf}
    \end{subfigure}
    \begin{subfigure}[b]{\linewidth}
        \captionsetup{justification=centering}
        \includegraphics[width=\linewidth]{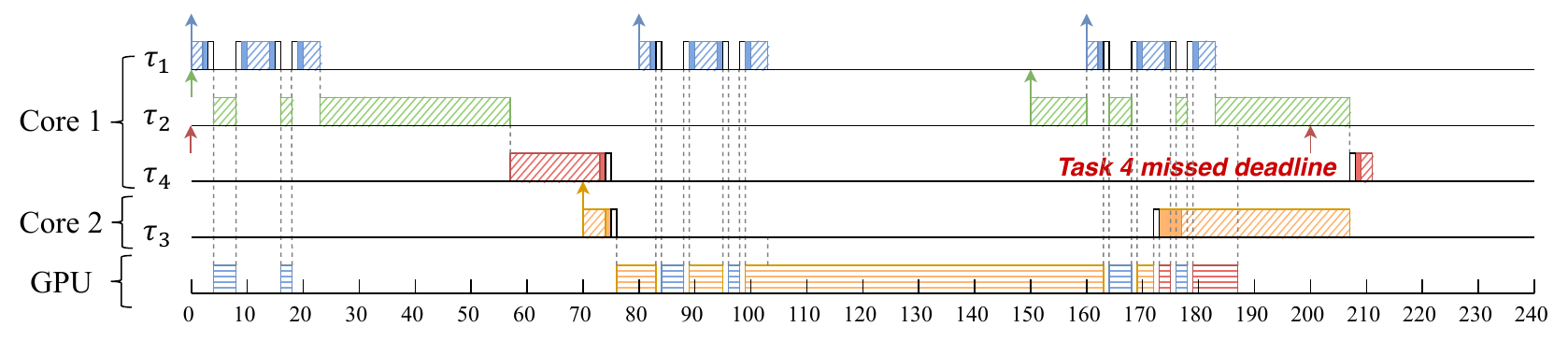}
        \vspace{-1.5\baselineskip}\caption{CPU priority = GPU priority: $\tau_1 > \tau_2 > \tau_3 > \tau_4$. RT test failed.}
        \label{fig:example_dft_prio}
    \end{subfigure}
    \begin{subfigure}[b]{\linewidth}
        \captionsetup{justification=centering}
        \includegraphics[width=\linewidth]{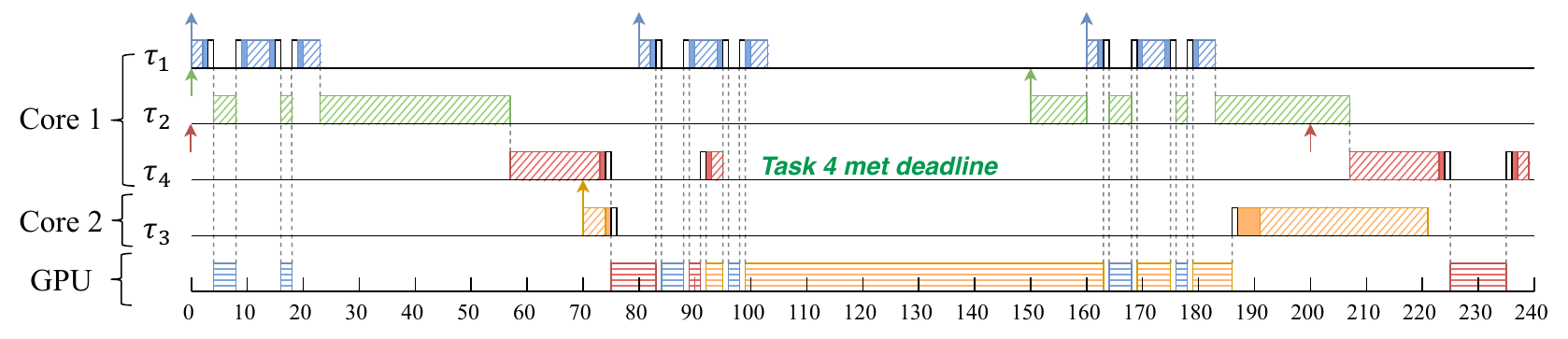}
        \vspace{-1.5\baselineskip}\caption{CPU priority: $\tau_1 > \tau_2 > \tau_3 > \tau_4$; GPU priority: $\tau_1 > \tau_2 > \tau_4 > \tau_3$. RT test passed.}
        \label{fig:example_sep_prio}
    \end{subfigure}
\caption{Example schedule of assigning separate GPU priority under self-suspension mode}
\label{fig:gpu_prio_example}
\end{figure}

Under the proposed approaches, GPU segments are executed following their OS-level task priorities by default, and the preemption can occur at segment boundaries. 

To improve taskset schedulability, we can assign separate priority to the GPU segments of a task, different from its CPU priority. 
In this case, a task $\tau_i$'s CPU segments and GPU segments may have different priorities, denoted as $\pi^c_i$ and $\pi^g_i$ respectively. Note that in our approach, the segments of the same type have the same priority (either $\pi^c_i$ or $\pi^g_i$).

We adopt Audsley's approach for this purpose~\cite{Audsley2007OPTIMALPA}. Hence, if the schedulability test given in Section~\ref{sec:analysis} determines a taskset is unschedulable, we iterate through all tasks from the lowest to the highest CPU priority and check whether each priority level can be assigned to the GPU segments of a task without causing the taskset to fail the schedulability test. Allowing different priorities for CPU and GPU segments may cause a deadlock if their priorities are not coordinated. To prevent deadlocks, we maintain the relative priority order of GPU segments identical to their corresponding CPU segments (i.e., OS-level priorities) for tasks executing on the same CPU core. For instance, consider two tasks $\tau_1$ and $\tau_2$ assigned to the same CPU, with CPU priority $\pi^c_1 > \pi^c_2$. If our algorithm suggests a GPU priority order where $\pi^g_1 < \pi^g_2$, we treat this assignment as infeasible.

\begin{example}[Effect of separate priority for GPU segment]
    Consider the taskset in Table~\ref{tab:taskset_separate_priority_example}. Task priority is assigned by the Rate-Monotonic (RM) policy, and CPUs are assigned by Worst-Fit-Decreasing (WFD) heuristic. The lower the task index, the higher its priority.
    The first job of $\tau_3$ arrives at time 70, and the other tasks' jobs arrive at time 0.
    In Fig.~\ref{fig:example_dft_prio}, each task uses the same priority for its CPU and GPU segments. During the time window $t=[73,74]$, both $G^e_{3,1}$ and $G^e_{4,1}$ were ready to begin. Since $G^e_{3,1}$ has a higher priority, it gained GPU access. When $G^e_{4,1}$ finished at $t=187$, $C_{4,2}$ could not start because $\tau_2$ is running on the same CPU. Those delays caused $\tau_4$ to miss the deadline.

    In Fig.~\ref{fig:example_sep_prio}, the priorities of tasks' CPU segments remain unchanged but we swapped the GPU priority of $\tau_3$ and $\tau_4$. Unlike the schedule in Fig.~\ref{fig:example_dft_prio}, at $t=75$, $G^e_{4,1}$ obtained GPU access first and $\tau_4$ is able to meet the deadline.

    This timeline illustrates only one possible scenario.  We evaluated this taskset using the proposed response time analysis detailed in Sec.~\ref{sec:analysis_preemptive}. The results show that the priority assignment in Fig.~\ref{fig:example_dft_prio} failed the test while the priority assignment in Fig.~\ref{fig:example_sep_prio} passed the test.
\label{example:sep_gpu_prio_suspend}
\end{example}

To implement this approach, we allow our macros to take one extra argument, which is the user-defined GPU segment priority: \texttt{gcapsGpuSegBegin(fd, getpid(), gprio)} and \texttt{gcapsGpuSegEnd(fd, getpid(), gprio)}.
In Alg.~\ref{alg:ioctl_approach},  we change line~\ref{alg2:tau_h_assigned_highest_prio}, line~\ref{alg2:comp_prio} and line~\ref{alg2:tau_k_assigned_highest_prio} to decide whether a task's TSG should be in the runlist or not based on tasks' GPU segment priorities, i.e., \texttt{gprio}.



\section{End-to-End Response Time Analysis}
\label{sec:analysis}
This section presents a comprehensive analysis of the end-to-end response time of tasks involving CPU and GPU interactions for both the round-robin approach of the default Nvidia Tegra driver and our proposed priority-based GPU context scheduling approach. The applicability of the proposed analysis is not limited to integrated GPUs with shared memory architecture such as Jetson series. As explained in Sec.~\ref{sec:gcaps_ctxsw_details}, only the runlist buffer is swapped during context switches and the GPU memory of a preempted task is not swapped out.

\subsection{Response Time Breakdown}\label{sec:response_time_breakdown}
Before proceeding to the individual analysis for each scheduling approach, we first present the components that account for the overall response time. For a task $\tau_i$, its worst-case response time can be upper-bounded by:
\begin{equation}
\small
\begin{aligned}
    R_i := C_i + G_i + I^C_i + I^G_i
\end{aligned}
\end{equation}
\smallskip
$C_i$ and $G_i$ stand for the computation requirement of $\tau_i$'s CPU and GPU segments. $I^C_i$ and $I^G_i$ are the interference $\tau_i$ can experience due to CPU segments and GPU segments, respectively.
The interference due to CPU segments, $I^C_i$, consists of two main components: (i) preemption from higher-priority tasks on the same core ($P^C_i$), and (ii) blocking due to runlist updates in our approach ($B^C_i$). These components will be explained later in Sec.~\ref{sec:analysis_preemptive}. The total CPU interference is expressed by:
\begin{equation}
\small
\begin{aligned}
    I^C_i := P^C_i + B^C_i
\end{aligned}
\end{equation}

\begin{figure}[t]
\vspace{-10pt}
\centering
    \begin{subfigure}[b]{0.33\linewidth}
        \captionsetup{justification=centering}
        \includegraphics[width=\linewidth]{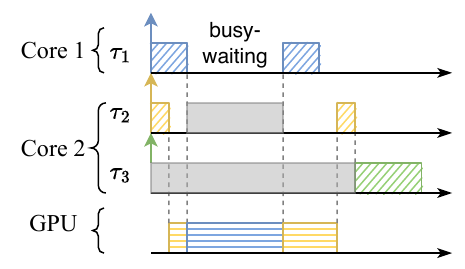}
        \vspace{-1\baselineskip}\caption{}
        \label{fig:gpu_preemption_dft_prio}
    \end{subfigure}\hfill
    \begin{subfigure}[b]{0.33\linewidth}
        \captionsetup{justification=centering}
        \includegraphics[width=\linewidth]{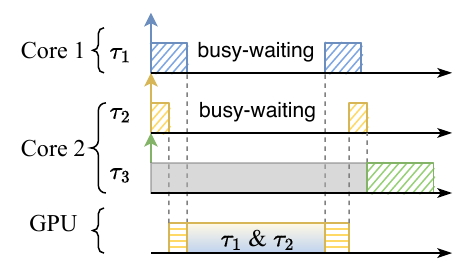}
        \vspace{-1\baselineskip}\caption{}
        \label{fig:rr_remote_delay}
    \end{subfigure}\hfill    
    \begin{subfigure}[b]{0.33\linewidth}
        \captionsetup{justification=centering}
        \includegraphics[width=\linewidth]{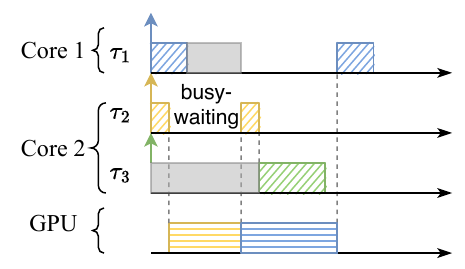}
        \vspace{-1\baselineskip}\caption{}
        \label{fig:gpu_preemption_sep_prio}
    \end{subfigure}
\caption{GPU Segment Interference with busy-waiting ($G^m$ and $\epsilon$ are omitted for simplicity). The shaded area indicates the amount of preemption or blocking. (a) Proposed preemptive scheduling: ($\pi^c_i=\pi^g_i$) $\pi_1 > \pi_2 > \pi_3$. (b) Default round-robin scheduling: $\tau_1$ indirectly delays $\tau_3$ regardless of the relative priority of $\tau_1$ and $\tau_3$. (c) Proposed preemptive scheduling with separate GPU segment priority assignment: $\pi^c_1 > \pi^c_2 > \pi^c_3$; $\pi^g_2 > \pi^g_1 > \pi^g_3$.}
\end{figure}

As for the interference from GPU segments, $I^G_i$, we break it down into the following components:

\noindent\textbf{Direct Preemption ($I^{dp}_i$). }
\textit{Direct preemption} occurs under the proposed approach when a task's GPU segment execution gets preempted by a higher-priority GPU segment, regardless of whether the two tasks are running on the same CPU core or not.
As an example, Fig.~\ref{fig:gpu_preemption_dft_prio} shows that $\tau_2$ gets direct preemption from $\tau_1$ under the proposed approach, since they are contending for the GPU resource.

\noindent\textbf{Indirect Delay ($I^{id}_i$).} 
\textit{Indirect delay} refers to the delay imposed on the CPU segments of a task $\tau_i$ due to busy-waiting GPU segments.\footnote{$I^{id}_i$ is not about the context switching overhead $\theta$ (Def.~\ref{def:ctxsw_overhead_theta}), which includes register saving/restoring cost and extra delay due to pixel-level or thread block-level preemption granularity as discussed in Sec.~\ref{sec:background}. The overhead $\theta$ (and $\epsilon=\alpha+\theta$ in our approach) will be discussed separately.} Such busy-waiting is caused by higher-priority tasks on the same CPU as $\tau_i$, and the amount of indirect delay imposed on $\tau_i$ is also affected by GPU-using tasks on different CPU cores.
However, indirect delay cannot exist stand-alone. It is contingent on the presence of direct interference (i.e. direct CPU or GPU preemption) from a higher-priority GPU-using ($\eta^g_h>0$) task $\tau_h$ running on the same core as $\tau_i$.

\begin{example}
[Indirect delay]
For the proposed approach, the CPU segment of $\tau_3$ in Fig.~\ref{fig:gpu_preemption_dft_prio} shows an example: $\tau_1$ preempts $\tau_2$’s GPU execution, making $\tau_2$’s busy waiting period longer, and it further delays $\tau_3$’s execution. In other words, $\tau_1$ indirectly delays $\tau_3$ since they are not directly competing for the same resource. For the default scheduling approach, consider the taskset in Fig.~\ref{fig:rr_remote_delay}, with priority of $\tau_2 > \tau_1 > \tau_3$. In this case, $\tau_3$ is delayed by the GPU interleaved execution of $\tau_1$ and $\tau_2$; thus $\tau_3$ is indirectly delayed by $\tau_1$ since $\tau_1$ and $\tau_3$ are not competing for the same resource. 
\end{example}

\noindent\textbf{Interleaved Execution ($I^{ie}_i$). }
The default round-robin scheduling approach adopts a scheduling strategy of \textit{interleaved execution} where multiple GPU kernels are executed in an alternating, overlapped manner, instead of completing one before starting the next, as discussed in Sec.~\ref{sec:background}. With this approach, a TSG of a task has to wait for the completion of the preceding other tasks' TSGs before it starts, leading to extended execution time of GPU segments, which are perceived as the prolonged boxes on the timeline in profiling tools.
In the worst case, all the TSGs are active and fully utilize the time slices, and each GPU segment of a task $\tau_i$ must wait for prior TSGs to finish. This procedure repeats throughout the entire execution of $\tau_i$. To model this delay, we consider a TSG time slice length of  $L$, and $G^e_{i,j}$ requires at most $\lceil\frac{G^e_{i,j}}{L}\rceil$ times of TSG slices and GPU context switches. Assuming there are $\nu$ GPU-using tasks in the system, the maximum delay for each GPU segment of $\tau_i$ due to interleaved execution is upper bounded by:
\begin{equation}
\begin{aligned}
    \mathcal{I}(\nu, G^e_{i,j}) := (L + \theta) \cdot \nu \cdot \lceil\frac{G^e_{i,j}}{L}\rceil
\end{aligned}
\label{eq:gpu_rr}
\end{equation}

By considering all the above factors, the overall GPU interference of a task $\tau_i$ can be computed as follows:
\begin{equation}
\small
\begin{aligned}
    I^G_i := I^{dp}_i + I^{id}_i + I^{ie}_i 
\end{aligned}
\end{equation}

\subsection{Analysis for Default Round-Robin TSG Scheduling}
\label{sec:analysis_tsg_rr}
In the prior literature, the tasks' response times are typically considered unpredictable under the default scheduling policy in the GPU driver (\cite{Elliott2012, Elliott_RTS13, Elliott_RTSS13, HKim2017, patel2018analytical}), and also the tasks' execution is graphed as large overlapped boxes in the GPU profiling tools. However, as we have revealed that the Tegra GPU driver employs a time-sliced round-robin scheduling approach in Sec.~\ref{sec:background}, this makes this approach analyzable.
In the following analysis, we assume that the default behavior of the Tegra driver that equally treats all GPU requests from different processes. This is a reasonable assumption given that the TSG slice and priority configurations are not exposed to the user and we have not observed changes in those settings, as we discussed before.

\begin{lemma}[GPU interleaved execution]
\label{lm:gpu_interleaved_rr}
    Under the default Tegra GPU driver, the worst-case interference from GPU interleaved execution for a task $\tau_i$ is bounded by:
\begin{equation}
\small
\begin{aligned}
    I^{ie}_i = \sum_{j=1}^{\eta^g_i} \mathcal{I}(|\{k \mid \tau_k \neq \tau_i \land \eta^g_k > 0\}|, G^e_{i,j} )
\end{aligned}
\end{equation}
\end{lemma}
\vspace{-10pt}
\begin{proof}
    This equation captures the total amount of interference due to the interleaved execution of any other tasks $\tau_k$ and $\tau_i$ itself.
\end{proof}

\begin{lemma}[GPU direct preemption]
    Under the default Tegra GPU driver, the GPU direct preemption delay imposed on a real-time task $\tau_i$ is zero, i.e., $I^{dp}_i=0$.
\end{lemma}
\vspace{-10pt}
\begin{proof}
    This follows our definition of direct preemption in Sec.~\ref{sec:response_time_breakdown}. With the default driver's approach, GPU segments are not preempted, but interleaved.
\end{proof}

\begin{lemma}[CPU blocking time]
    Under the default Tegra GPU driver, the worst-case CPU blocking time for a task $\tau_i$ is zero, i.e., $B^C_i=0$.
\end{lemma}
\vspace{-10pt}
\begin{proof}
    This is obvious since the default driver does not require tasks to explicitly request for the runlist update.
\end{proof}

\subsubsection{Busy-Waiting Mode}

\begin{lemma}[GPU indirect delay]
\label{lm:gpu_indirect_rr_busy}
Under the default Tegra GPU driver with busy-waiting, the worst-case interference from GPU indirect delay for a task $\tau_i$ is bounded by:
\begin{equation}
\small
\begin{aligned}
    I^{id}_i &= \sum_{\tau_h\in hpp(\tau_i) \land \eta^g_h>0 } \lceil\frac{R_i}{T_h}\rceil \cdot \sum_{j=1}^{\eta^g_{h}} \mathcal{I}(| \{ k \mid \tau_{k} \notin hpp(\tau_i) \land \eta^g_{k}>0 \cup \tau_h \} |, G^e_{h,j})
\end{aligned}
\end{equation}
where $hpp(\tau_i)$ is the set of higher-priority tasks running on the same CPU core as $\tau_i$.
\end{lemma}
\vspace{-10pt}
\begin{proof}
    A task $\tau_i$ experiences indirect delay from each higher-priority task $\tau_h$ busy waiting on the same CPU during its GPU segment execution ($\tau_h\in hpp(\tau_i) \land \eta^g_h>0$ in the outer summation). 
    For each $\tau_h$, we need to bound the maximum busy-waiting period during which its GPU execution can interleave with any other GPU-using tasks. This can be done by the summation of Eq.~\eqref{eq:gpu_rr} for each GPU segment of $\tau_h$,  $G_{h,j}^e$. However, since the busy-waiting periods of tasks in $hpp(\tau_i)$, other than $\tau_h$, have already been accounted for iteratively in the outer summation, we exclude $hpp(\tau_i)$ from the cardinality of the interleaving taskset considered by Eq.~\eqref{eq:gpu_rr}, i.e., $|\{k \mid\tau_{k} \notin hpp(\tau_i) \land \eta^g_{k}>0 \cup \tau_h\}|$, to prevent double counting.    
\end{proof}

\begin{lemma}[CPU preemption]
\label{lm:cpu_preemption_rr_busy}
    Under the default Tegra driver with busy-waiting, the worst-case interference from CPU preemption is bounded by:
    \begin{equation}
    \small
    \begin{aligned}
        P^C_i = \sum_{\tau_h \in hpp(\tau_i)} \lceil\frac{R_i}{T_h}\rceil \cdot (C_h + G^m_h)
    \end{aligned}
    \end{equation}
\end{lemma}
\vspace{-10pt}
\begin{proof}
    A task $\tau_i$ can only experience interference from CPU preemption from the higher-priority tasks running on the same CPU core for a duration of $C_h+G^m_h$.
\end{proof}



\subsubsection{Self-Suspension Mode}\label{sec:analysis_rr_suspend}

\begin{lemma}[GPU indirect delay]
\label{lm:gpu_indirect_rr_suspend}
    Under the default Tegra GPU driver with self-suspension, the worst-case interference from GPU indirect delay for a task $\tau_i$ is zero, i.e., $I^{id}_i=0$.
\end{lemma}
\vspace{-10pt}
\begin{proof}
    A task with self-suspension would not experience GPU indirect delay as explained in Sec.~\ref{sec:response_time_breakdown}.
\end{proof}

\begin{lemma}[CPU preemption]
\label{lm:cpu_preemption_rr_suspend}
    Under the default Tegra driver self-suspension, the worst-case interference from CPU preemption is bounded by:
    \begin{equation}
    \small
    \begin{aligned}
        P^C_i = \sum_{\tau_h \in hpp(\tau_i)} \lceil\frac{R_i+J^c_h}{T_h}\rceil \cdot (C_h + G^m_h)
    \end{aligned}
    \end{equation}
    where $J^c_h = R_h - (C_h + G^m_h)$.
\end{lemma}
\vspace{-10pt}
\begin{proof}
    Each job of $\tau_h$ imposes a delay of up to $C_h + G^m_h$ on $\tau_i$ and the self-suspending behavior of $\tau_h$ can be captured by a jitter term $J^c_h$ as reported in~\cite{Bletsas2018}.
\end{proof}



\subsection{Analysis for Proposed GPU Context Scheduling}
\label{sec:analysis_preemptive}
Our approach introduces the runlist update delay of $\epsilon$ (Def.~\ref{def:runlist_update_delay}). In the worst case, runlist updates are required both before and after each GPU segment of $\tau_i$, since the associated TSGs need to be added and removed by IOCTL calls as shown in Listing~\ref{lst:ioctl_sample}. This leads to a cumulative cost of $2\epsilon \cdot \eta^g_i$ for the entire job of $\tau_i$.
For ease of presentation, we define $G_i^*$, $G_i^{e*}$, and $G_i^{m*}$ to incorporate the two times of runlist updates into the execution requirements.
\begin{equation*}
\small
\begin{aligned}
    G^*_i=G_i + 2\epsilon \cdot \eta^g_i\; \text{,}\; G_i^{e*} = G^e_i+2\epsilon\cdot\eta^g_i     
\;\text{and}\; G_i^{m*} = G^m_i+2\epsilon\cdot\eta^g_i
\end{aligned}
\end{equation*}

Also, due to the use of \texttt{rt-mutex} in our approach, each GPU segment of $\tau_i$ can experience blocking time of $\epsilon$ when there is an ongoing runlist update initiated by a lower-priority task.

\begin{figure}[]
\vspace{-10pt}
\centering
    \includegraphics[width=\linewidth]{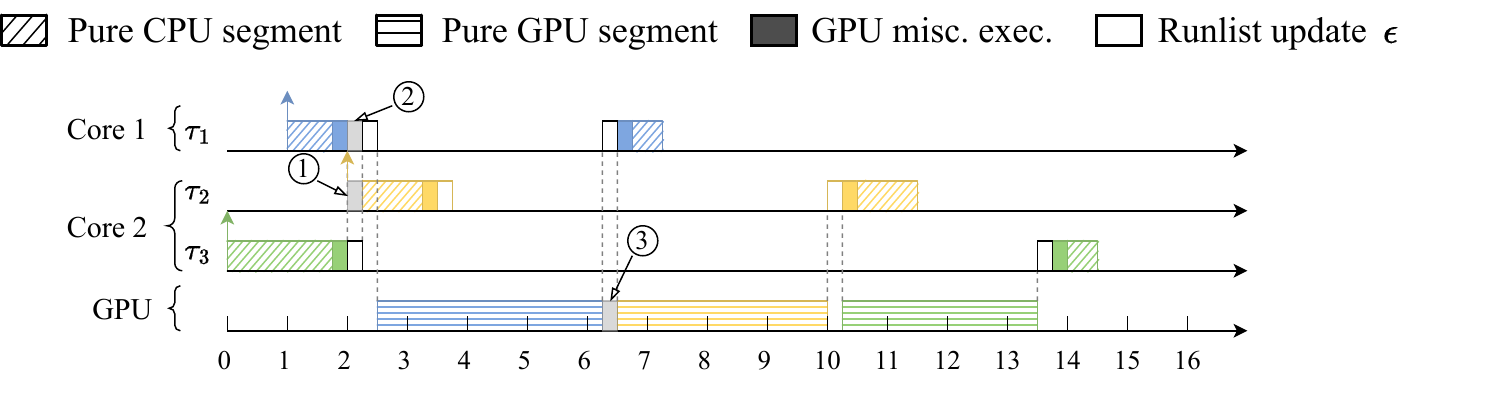}
   \vspace{-1\baselineskip}\caption{Example schedule of three tasks ($\pi_1 > \pi_2 > \pi_3$) with runlist update delay}
\label{fig:ioctl_delay}
\end{figure}

\begin{lemma}[Blocking time]
\label{lm:rl_blocking_proposed}
    Under the proposed approach, the worst-case blocking time for a task $\tau_i$ is bounded by:
    \begin{equation}
    \begin{aligned}
        B^C_i = (\eta^g_i + 1) \cdot \epsilon
    \end{aligned}
    \end{equation}
\end{lemma}
\vspace{-10pt}
\begin{proof}
    First, at least one time of the blocking of $\epsilon$ applies to every $\tau_i$, regardless of whether it is a GPU-using or CPU-only task. This is because it can experience blocking from GPU-using lower-priority tasks at the very beginning of its job instance, as illustrated by \ding{192} in Fig.~\ref{fig:ioctl_delay}. If $\tau_i$ is a GPU-using task, each GPU segment requires up to $\epsilon$ for potential blocking from lower-priority tasks, which results in $\eta^g_h\cdot\epsilon$. Therefore, the total amount of blocking imposed by lower-priority tasks is bounded by $(\eta^g_i + 1) \cdot \epsilon$.
\end{proof}


\begin{lemma}[Interleaved execution]
    Under the proposed approach, the interference from interleaved execution of a real-time task $\tau_i$ is zero, i.e., $I^{ie}_i=0$.
\end{lemma}
\vspace{-10pt}
\begin{proof}
    This is obvious since the GPU segments of real-time tasks are not allowed to execute in an interleaved manner with our proposed approach.
\end{proof}

\begin{example}[Runlist update delay]
Fig.~\ref{fig:ioctl_delay} can help better understand all types of runlist update delay under the proposed approach. The task of interest here is $\tau_2$ which runs with medium priority.
$\tau_3$ triggers runlist update first, and this blocks to $\tau_2$'s CPU segment at \ding{192} and $\tau_1$'s runlist update at \ding{193}.  
Before $\tau_1$ finishes GPU execution, $\tau_2$ cannot start its GPU segment as $\tau_1$ is actively using the GPU with higher priority than $\tau_2$. Then, the start time of $\tau_2$'s GPU segment is  delayed by the runlist update at \ding{194} triggered by $\tau_1$ to remove $\tau_1$'s TSG from the runlist. 
\end{example}
Based on these observations, we derive the response time analysis of busy-waiting and self-suspending GPU tasks in the following.

\subsubsection{Busy-Waiting Mode}

\begin{lemma}[GPU direct preemption]
\label{lm:gpu_direct_proposed_busy}
    Under the proposed approach with busy-waiting, the worst-case interference from GPU direct preemption for a task $\tau_i$ is bounded by:
\begin{equation}
\small
\begin{aligned}
    I^{dp}_i &= \sum_{\substack{\tau_h \in hpp(\tau_i) \\\land \eta^g_h>0 \land \eta^g_i>0}} \lceil \frac{R_i}{T_h} \rceil \cdot G^{e*}_h 
    + \sum_{\substack{\tau_h\in hp(\tau_i) \land \tau_h\notin hpp(\tau_i) \\ \land\eta^g_h>0 \land \eta^g_i>0 }} \lceil \frac{R_i+J^g_h}{T_h} \rceil \cdot G_h^{e*}
\end{aligned}
\label{eq:gpu_direct_proposed_busy}
\end{equation}
where $J^g_h=R_h-G^e_h$.
\end{lemma}
\vspace{-10pt}
\begin{proof}
A task $\tau_i$ can experience direct GPU preemption from a higher-priority task $\tau_h$ only when both $\tau_i$ and $\tau_h$ are GPU-using tasks ($\eta^g_i>0 \land \eta^g_h>0$).
We consider the preemption from: (i) $\tau_h$ on the same core as $\tau_i$ ($\tau_h \in hpp(\tau_i)$), and (ii) on different cores ($\tau_h \in hp(\tau_i) \land \tau_h \notin hpp(\tau_i)$). 
In both of the cases, the duration of preemption equals to $G^{e*}_h$, which includes the pure GPU execution and the runlist update cost of $2\epsilon$ for each of $\tau_h$'s GPU segment. 

In the case of (ii), a release jitter $J^g_h$ is considered to account for the carry-in effect of $\tau_h$'s GPU execution. This is because $\tau_h$'s job releases are not synchronized with the $\tau_i$'s job release, causing carry-in jobs to the window of $\tau_i$'s response time. It is known that in an arbitrary time window $t$, the number of arrivals of a higher-priority task $\tau_h$ with a carry-in job can be upper-bounded by $\lceil\frac{t+J_h}{T_h}\rceil$, where the jitter $J_h=D_h-C_h$ if $\tau_h$'s response time is unknown~\cite{bertogna2008schedulability}, and $J_h=R_h-C_h$ otherwise~\cite{lee2017improved}.
\end{proof}

\begin{lemma}[GPU indirect delay]
\label{lm:gpu_indirect_proposed_busy}
Under the proposed approach with busy-waiting, the worst-case interference from GPU indirect delay for a task $\tau_i$ is bounded by:
\begin{equation}
\small
\begin{aligned}
    I^{id}_i = \sum_{\substack{\tau_h\in hp(\tau_i) \land \tau_h\notin hpp(\tau_i) \\ \land\eta^g_h>0 \land \eta^g_i=0 }} \lceil \frac{R_i+J^g_h}{T_h} \rceil \cdot G_h^{e*}
\end{aligned}
\end{equation}
\end{lemma}
\vspace{-10pt}
\begin{proof}
Under the proposed approach with busy-waiting, a task $\tau_i$ gets GPU indirect delay when it is preempted by a higher-priority task $\tau_h$ on the same core when $\tau_h$ is busy waiting on the CPU during its GPU segment execution. During this busy-waiting period, the higher-priority task $\tau_h$ can experience GPU direct preemption from any higher-priority GPU-using tasks on different CPUs ($\tau_h\in hp(\tau_i) \land \tau_h\notin hpp(\tau_i) \land \eta^g_h>0$), and such $\tau_h$ further increases indirect delay imposed on $\tau_i$.
This happens regardless of whether $\tau_i$ is a CPU-only or GPU-using task. However, the last term in Eq.~\eqref{eq:gpu_direct_proposed_busy} has already bounded all the delays from such $\tau_h$ that $\tau_i$ may experience when $\tau_i$ is a GPU-using task. Therefore, in this lemma, we only consider the case that $\tau_i$ is a CPU-only task, to prevent double counting.
\end{proof}

\begin{lemma}[CPU preemption]
\label{lm:cpu_preemption_proposed_busy}
    Under the proposed approach with busy-waiting, the worst-case interference from CPU preemption of a task $\tau_i$ is bounded by:
    \begin{equation}
    \small
    \begin{aligned}
        P^C_i = \sum_{\tau_h \in hpp(\tau_i)} \lceil \frac{R_i}{T_h} \rceil \cdot (C_h + G^{m}_h)
    \end{aligned}
    \end{equation}
\end{lemma}
\vspace{-10pt}
\begin{proof}
    The proof directly follows Lemma~\ref{lm:cpu_preemption_rr_busy}.
\end{proof}


\subsubsection{Self-Suspension Mode}

\begin{lemma}[GPU direct preemption]
\label{lm:gpu_direct_proposed_suspend}
    Under the proposed approach with self-suspension, the worst-case interference from GPU direct preemption for a task $\tau_i$ is bounded by:
\begin{equation}
\small
\begin{aligned}
    I^{dp}_i = \sum_{\substack{\tau_h\in hpp(\tau_i) \\\land \eta^g_h>0 \land \eta^g_i>0}} \quad \lceil \frac{R_i+J^g_h}{T_h} \rceil \cdot G^e_h 
        + \sum_{\substack{\tau_h\in hp(\tau_i)\land\tau_h\notin hpp(\tau_i) \\\land \eta^g_h>0 \land \eta^g_i>0}} \quad \lceil \frac{R_i+J^g_h}{T_h} \rceil \cdot G_h^{e*} 
\end{aligned}
\end{equation}
\end{lemma}
\vspace{-10pt}
\begin{proof}
This is a variant of Lemma~\ref{lm:gpu_direct_proposed_busy}. The difference lies in the first term accounting for the interference of GPU direct preemption from higher-priority GPU-using task $\tau_h$ on the same core. 
$\tau_h$ imposes interference of its pure GPU execution, $G_h^{e*}$, to $\tau_i$ only when $\tau_i$ uses the GPU ($\eta_i^g>0$). Here, from the perspective of $\tau_i$ with $\eta_i^g>0$, the runlist update delay on the CPU and GPU overlaps, and thus using $G_h^e$ safely bounds GPU preemption from $\tau_h$.
The self-suspending behavior of $\tau_h$ can be captured by a jitter term, as reported in~\cite{Bletsas2018}. The second term remains the same as in Lemma~\ref{lm:gpu_direct_proposed_busy}.
\end{proof}

\begin{lemma}[GPU indirect delay]
\label{lm:gpu_indirect_proposed_suspend}
    Under the proposed approach with self-suspension, the worst-case interference from GPU indirect delay of a task $\tau_i$ is zero, i.e., $I^{id}_i=0$.
\end{lemma}
\vspace{-10pt}
\begin{proof}
    This phenomena does not exist under self-suspension mode as explained in Sec.~\ref{sec:response_time_breakdown}.    
\end{proof}

\begin{lemma}[CPU preemption]
\label{lm:cpu_preeemption_proposed_suspend}
    Under the proposed approach with self-suspension, the worst-case interference from CPU preemption of a task $\tau_i$ is bounded by:
\begin{equation}
\small
\begin{aligned}
    P^C_i =
    \sum_{\substack{\tau_h \in hpp(\tau_i) \\\land \eta^g_h=0}} \lceil \frac{R_i}{T_h} \rceil \cdot C_h 
    + \sum_{\substack{\tau_h \in hpp(\tau_i) \\\land \eta^g_h>0}} \lceil \frac{R_i + J^c_h}{T_h} \rceil \cdot (C_h + G^{m*}_h)
\end{aligned}
\end{equation}
\end{lemma}
\vspace{-10pt}
\begin{proof}
    This is a variant of Lemma~\ref{lm:cpu_preemption_proposed_busy}. If $\tau_h$ is a GPU-using task running on the same core ($\tau_h \in hpp(\tau_i) \land \eta^g_h>0$), each job of $\tau_h$ imposes a delay of up to $(C_h + G^{m*}_h)$ and the self-suspending behavior of $\tau_h$ is accounted for by the jitter term, $J^c_h$~\cite{Bletsas2018}.
\end{proof}


\subsection{Analysis for GPU Priority Assignment}
\label{sec:analysis_gpu_prio}
When the GPU priority assignment given in Sec.~\ref{sec:separate_gpu_prio} is used, the amount of preemption due to higher-priority GPU tasks, i.e., $hpp()$ and $hp()$, needs to be revisited. 
Recall that our assignment preserves the same relative priority order for GPU segments as the CPU priority order for tasks running on the same core. The meaning of $hpp()$ therefore remains unchanged. However, $hp()$ needs to be redefined such that it means the set of tasks with higher ``GPU segment'' priorities in the system. This is because any interference due to GPU segment execution ($I^{dp}_i$ and $I^{id}_i$) of Lemma~\ref{lm:gpu_direct_proposed_busy},~\ref{lm:gpu_indirect_proposed_busy} and~\ref{lm:gpu_direct_proposed_suspend} are now governed by GPU segment priorities. When computing the release jitter $J^x_h$, $R_h$ needs to be replaced with $D_h$ since the worst-case response time of higher-priority tasks is unknown when applying our GPU priority assignment method. With these simple modifications, our analysis in the previous section can analyze the effect of the GPU priority assignment.

Besides the benefits introduced in Example~\ref{example:sep_gpu_prio_suspend}, the use of separate GPU segment priority assignment is particularly effective in mitigating the scheduling inefficiency of busy-waiting as shown in Example~\ref{example:sep_gpu_prio_busy}. Our evaluation results in Sec.~\ref{sec:schedulability_experiments} will confirm this claim. 
\begin{example}[Separate GPU priority assignment under busy-waiting mode]
\label{example:sep_gpu_prio_busy}
In Fig.~\ref{fig:gpu_preemption_dft_prio}, $\tau_3$ is indirectly preempted by $\tau_1$ as explained in Sec.~\ref{sec:response_time_breakdown}. However, if we consider Fig.~\ref{fig:gpu_preemption_sep_prio} where $\tau_2$ is assigned a higher GPU priority than $\tau_1$, $\tau_3$ no longer experiences the delay from $\tau_1$'s GPU segment, thereby achieving a shorter response time.
\end{example}

\section{Evaluation}
We conduct schedulability experiments to compare the proposed approaches against prior work and assess the effect of the GPU priority assignment. Then, we present a case study on two Nvidia embedded platforms.

\subsection{Schedulability Experiments}
\label{sec:schedulability_experiments}
We generated 1,000 random tasksets for each experimental setting based on the parameters in Table~\ref{tab:params_for_taskset_generation}. The parameter selection is inspired by the prior work~\cite{patel2018analytical}, with slight modifications to increase the system load. Based on the measurement in Sec.~\ref{sec:system_eval}, we aggressively set $\epsilon$ to 1~ms for our approaches, while assuming zero overhead for synchronization-based approaches and setting $\theta$ as low as 200$\mu$s for TSG context switching in the default round-robin scheduling.
For each task in a taskset, the number of tasks on each CPU is first chosen randomly within the range, and the utilization per CPU is generated based on the UUniFast algorithm~\cite{uunifast}. Then for each task, its period and the number of GPU segments are uniformly randomized within the given range. Then the parameters for each segment are determined. Task priority is assigned by the Rate Monotonic (RM) policy. After this, we re-allocate the tasks to the CPUs for load balancing purpose with WFD (worst-fit decreasing) heuristic. 

\begin{figure}[t]
\vspace{-10pt}
    \begin{subfigure}[t]{\linewidth}
    \vspace{-15pt}
        \includegraphics[width=\linewidth]{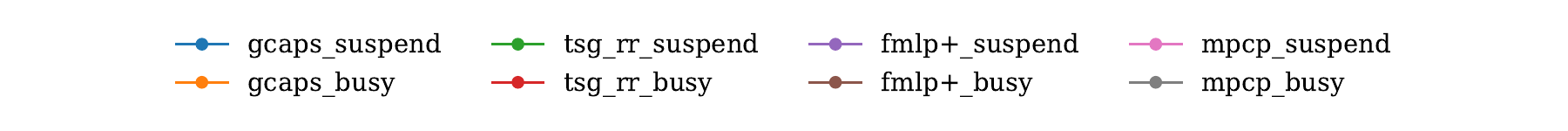}
    \end{subfigure}
    
\begin{minipage}{0.04\linewidth}
\hspace{-10pt}
    \begin{subfigure}[t]{\linewidth}
    \vspace{-1\baselineskip}
        \includegraphics[width=\linewidth]{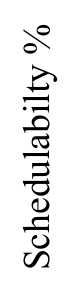}
    \end{subfigure}
\end{minipage}
\begin{minipage}{1\linewidth}
    \hspace{-16pt}
    \begin{subfigure}[t]{0.33\linewidth}
        \vspace{-0.5\baselineskip}
        \captionsetup{justification=centering}
        \includegraphics[width=\linewidth]{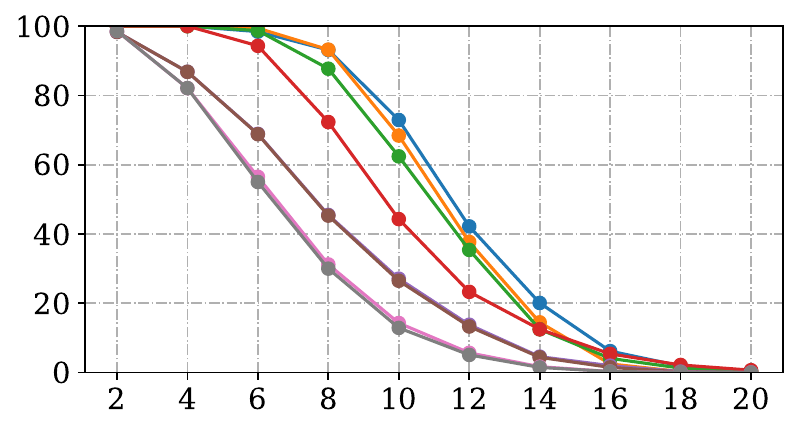}
        \vspace{-1.5\baselineskip}\caption{Number of tasks}
        \label{fig:grp1:num_of_tasks}
    \end{subfigure}\hspace{-5pt}
    \begin{subfigure}[t]{0.33\linewidth}
        \vspace{-0.5\baselineskip}
        \captionsetup{justification=centering}
        \includegraphics[width=\linewidth]{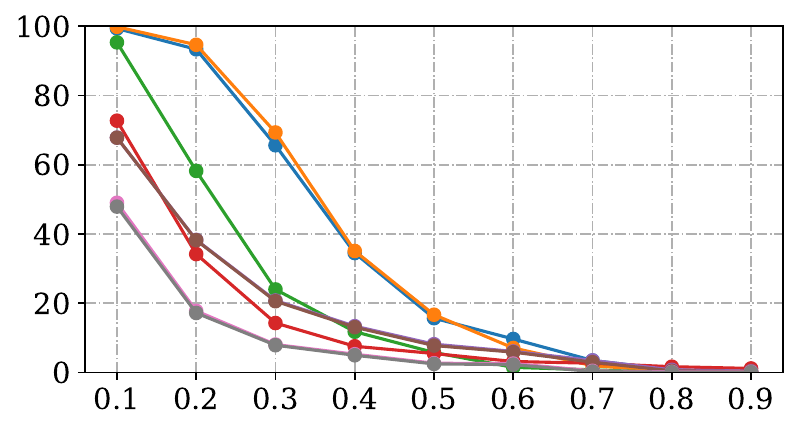}
        \vspace{-1.5\baselineskip}\caption{Utilization per CPU}
        \label{fig:grp1:util_per_cpu}
    \end{subfigure}\hspace{-5pt}
    \begin{subfigure}[t]{0.33\linewidth}
    \vspace{-0.5\baselineskip}
        \captionsetup{justification=centering}
        \includegraphics[width=\linewidth]{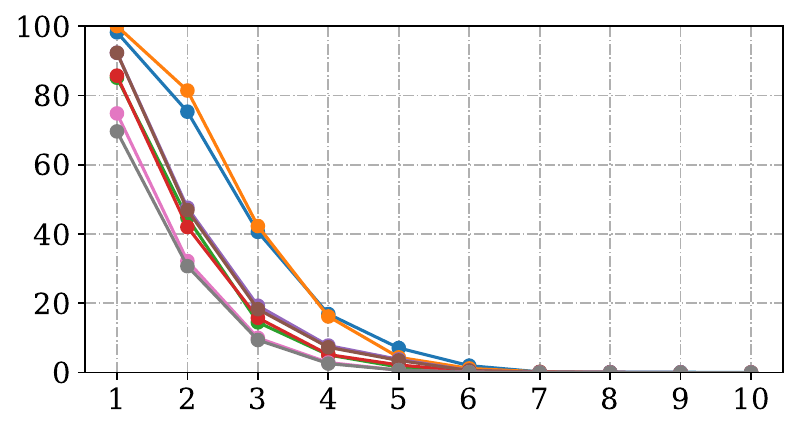}
        \vspace{-1.5\baselineskip}\caption{Number of CPUs}
        \label{fig:grp1:num_of_cpus}
    \end{subfigure}
\end{minipage}

\begin{minipage}{0.04\linewidth}
\hspace{-10pt}
    \begin{subfigure}[t]{\linewidth}
    \vspace{-1\baselineskip}
        \includegraphics[width=\linewidth]{figs/results/sim/general/schedulability_ylabel.pdf}
    \end{subfigure}
\end{minipage}
\begin{minipage}{1\linewidth}
\hspace{-16pt}
    \begin{subfigure}[t]{0.33\linewidth}
        \captionsetup{justification=centering}
        \includegraphics[width=\linewidth]{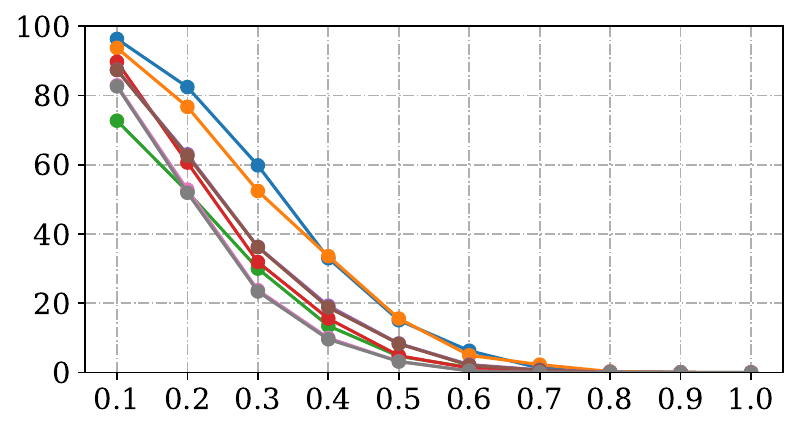}
        \vspace{-1.5\baselineskip}\caption{Ratio of GPU-using tasks}
        \label{fig:grp1:ratio_of_gpu_tasks}
    \end{subfigure}\hspace{-5pt}
    \begin{subfigure}[t]{0.33\linewidth}
        \captionsetup{justification=centering}
        \includegraphics[width=\linewidth]{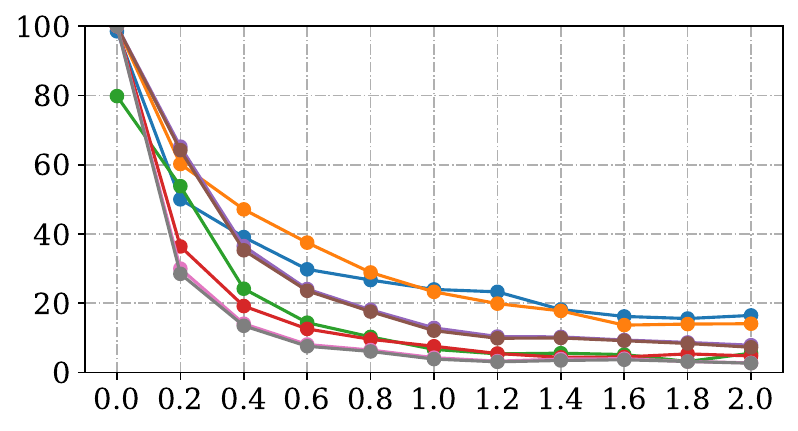}
        \vspace{-1.5\baselineskip}\caption{Ratio of $G_i$ to $C_i$}
        \label{fig:grp1:ratio_of_g_to_c}
    \end{subfigure}\hspace{-5pt}
    \begin{subfigure}[t]{0.33\linewidth}
        \captionsetup{justification=centering}
        \includegraphics[width=\linewidth]{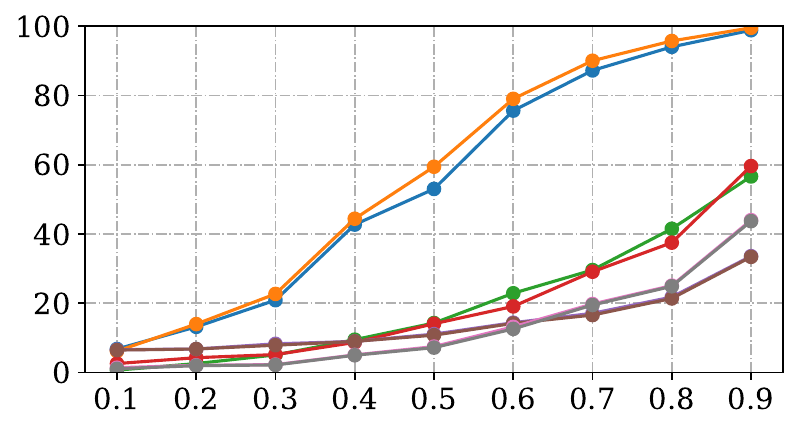}
        \vspace{-1.5\baselineskip}\caption{Ratio of best-effort tasks}
        \label{fig:grp1:ratio_of_be_tasks}
    \end{subfigure}
\end{minipage}

\caption{Schedulability of approaches with different experimental settings}
\end{figure}

\begin{table}[b]
\small
\centering
\begin{tabular}{l|l}
\hline
\textbf{Parameters}                                & \textbf{Value}       \\ \hline
Number of CPUs                                     & 4           \\
Number of tasks per CPU                            & [3, 6]      \\
Ratio of GPU-using tasks                           & [40, 60] \%   \\
Utilization per CPU                                  & [0.4, 0.6] \\
Task Period                                        & [30, 500] ms \\
Number of GPU segments per task                    & [1, 3]      \\
Ratio of GPU exec. to CPU exec. ($G_i/C_i$)        & [0.2, 2]    \\
Ratio of GPU misc. in GPU exec. ($G^m_i/G_i$)      & [0.1, 0.3]  \\
Runlist update cost ($\epsilon$)               & 1 ms  \\ \hline
\end{tabular}
\caption{Parameters for taskset generation}
\label{tab:params_for_taskset_generation}
\end{table}

\subsubsection{Comparison with Prior Work}

We first compare our proposed approach, GCAPS, with the default TSG round-robin scheduling as well as two well-known synchronization-based methods, MPCP~\cite{patel2018analytical} and FMLP+~\cite{brandenburg2014fmlp+}, both of which offer suspension-aware and busy-waiting analyses. For default TSG scheduling, we use the analysis in Sec.~\ref{sec:analysis_tsg_rr} and set the length of the time slice to 1024 $\mu$s since it is the default value set in the driver. For our approach, we use the analysis with the GPU priority assignment in Sec.~\ref{sec:separate_gpu_prio}. Hence, we first run the response time test for a taskset with the default RM priorities, and if the test fails, try again with separate priorities for GPU segments.

We investigate the impact of varying the number of tasks in the taskset, the number of CPUs, the utilization per CPU, and the ratio of GPU-using tasks in Figs.~\ref{fig:grp1:num_of_tasks},~\ref{fig:grp1:num_of_cpus}, ~\ref{fig:grp1:util_per_cpu} and~\ref{fig:grp1:ratio_of_gpu_tasks}, respectively. 
The results show that, in general, the \texttt{gcaps\_busy} and \texttt{gcaps\_suspend} approaches outperform previous methods. 

Fig.~\ref{fig:grp1:ratio_of_g_to_c} examines the effect of changing the ratio of $G_i/C_i$. When the ratio of $G_i/C_i$ is small, the proposed approach underperforms \texttt{fmlp+} and \texttt{tsg\_rr\_suspend}, because \texttt{fmlp+} can efficiently schedule tasks when GPU load is light, and the delay caused by TSG context switching is not significant when the duration of GPU segments are relatively short. The advantages of our approaches are mitigated by the critical section of runlist updates, but this trend does not continue as the ratio increases.

Lastly, we explore the impact of best-effort tasks running with the lowest priority in the system.
After generating the tasks using the aforementioned method, we randomly designate a specific percentage of tasks as best-effort tasks in this experiment.
Fig.~\ref{fig:grp1:ratio_of_be_tasks} depicts the percentage of schedulable tasksets as the ratio of best-effort tasks increases. The rest of the tasks are all real-time tasks with constraint deadlines. The best-effort tasks contribute to blocking time in the analysis of \texttt{mpcp} and \texttt{fmlp+}, and share the time-sliced GPU with real-time tasks in \texttt{tsg\_rr}. Since GPU preemption is enabled in our proposed approaches, they significantly outperform the prior methods.

\subsubsection{Effect of GPU Priority Assignment}
\begin{figure}[t]
\vspace{-10pt}
\centering
    \includegraphics[width=\linewidth]{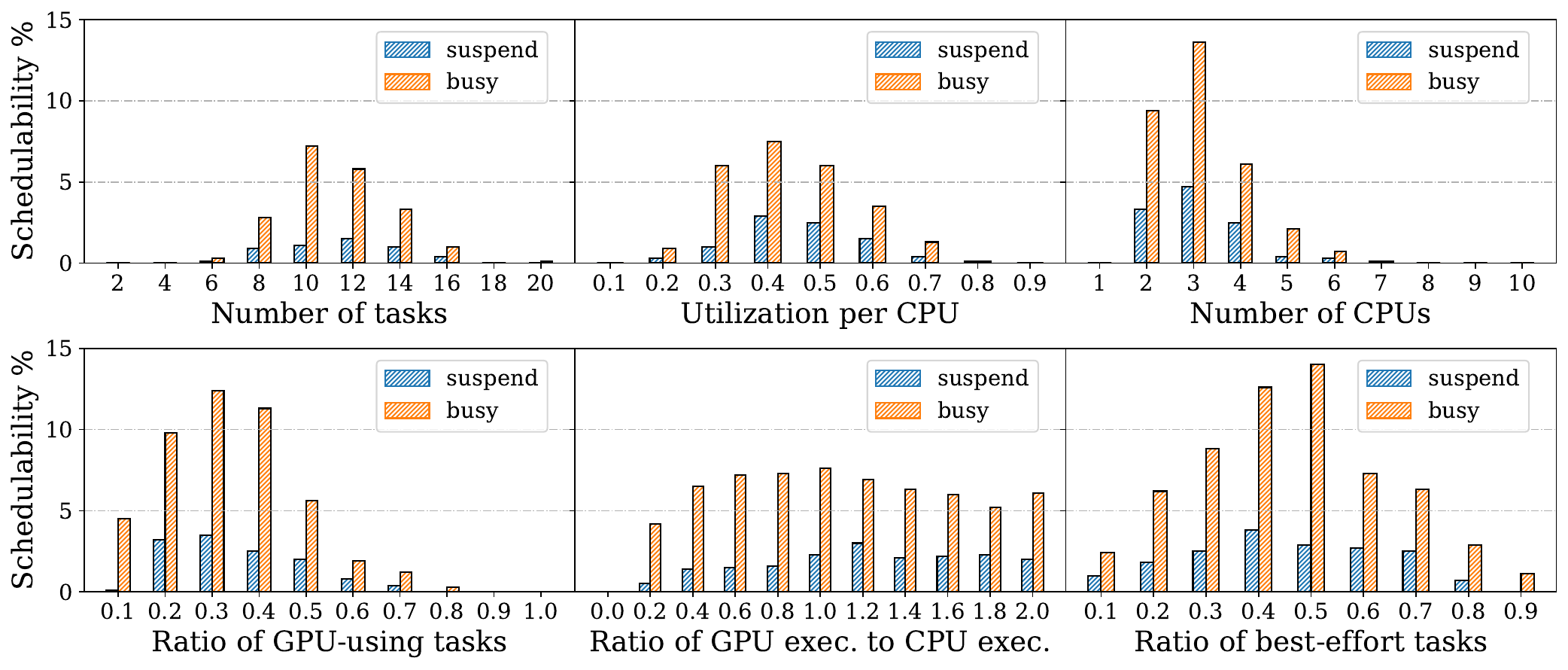}
\caption{Schedulability gain in GCAPS by GPU priority assignment}
\label{fig:gain_gpu_prio}
\end{figure}

In this experiment, we evaluate the impact of GPU priority assignment on taskset schedulability. We compare baseline analyses of \texttt{gcaps\_busy} and \texttt{gcaps\_suspend} with and without separate GPU priorities, using the taskset generation parameters from Table~\ref{tab:params_for_taskset_generation}.  
Fig.~\ref{fig:gain_gpu_prio} illustrates the advantages of GPU priority assignment. Busy-waiting approaches tend to benefit more from this assignment, as explained in Section~\ref{sec:analysis_gpu_prio} of the manuscript. 
Additionally, both busy-waiting and self-suspending approaches benefit from it since assigning GPU priorities independently of CPU priorities makes GPU resource allocation more efficient. E.g. Tasks with shorter GPU segments or higher GPU urgency can be prioritized appropriately, reducing resource wastage.


\subsection{System Evaluation}
\label{sec:system_eval}

We implemented our preemptive GPU scheduling approaches on two Nvidia platforms: the Nvidia Jetson Xavier NX Development Kit running L4T R35.2.1 with Jetpack 5.0.2 and the Nvidia Jetson Orin Nano Developer Kit running L4T R35.4.1 with Jetpack 5.1.2. The first platform features a 6-core 64-bit Carmel ARMv8.2 processor and a Volta architecture GPU. For our experiments, we configured it to run at its highest frequencies in the 6-core 15W mode. The second platform is equipped with a 6-core Arm Cortex-A78AE v8.2 64-bit CPU and an Ampere architecture GPU, and we operated it at its peak frequencies under its default power mode.

\begin{table}[b]
\small
\centering
\begin{tabular}{cc|ccccc}
\hline
Task & Workload        & $C_i$ & $G_i$ & $T_i=D_i$ & CPU & Priority \\ \hline
1    & histogram       & 1    & 10   & 100       & 1   & 70       \\
2    & mmul\_gpu\_1    & 2    & 12   & 150       & 2   & 69       \\
3    & mmul\_cpu    & 67   & 0    & 200       & 2   & 68       \\
4    & projection            & 12   & 15   & 300       & 1   & 67       \\
5    & dxtc            & 2    & 16   & 400       & 1   & 66       \\
6    & mmul\_gpu\_2    & 4    & 44   & 200       & 4   & 0        \\
7    & simpleTexture3D (graphic app) & 4    & 27   & 67        & 4,5   & 0        \\ \hline
\end{tabular}
\caption{Taskset used in case study}
\label{tab:taskset_case_study_nx}
\end{table}


\begin{figure}[t]
\vspace{-10pt}
    \centering
    \begin{subfigure}[b]{0.9\linewidth}
    \vspace{-10pt}
        \includegraphics[width=\linewidth]{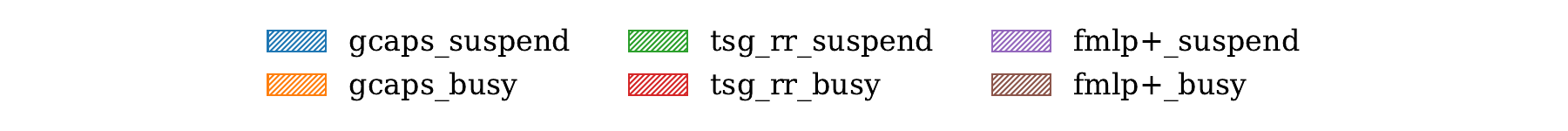}
    \end{subfigure}
    \begin{subfigure}[b]{0.49\linewidth}
        \vspace{-0.5\baselineskip}
        \includegraphics[width=\linewidth]{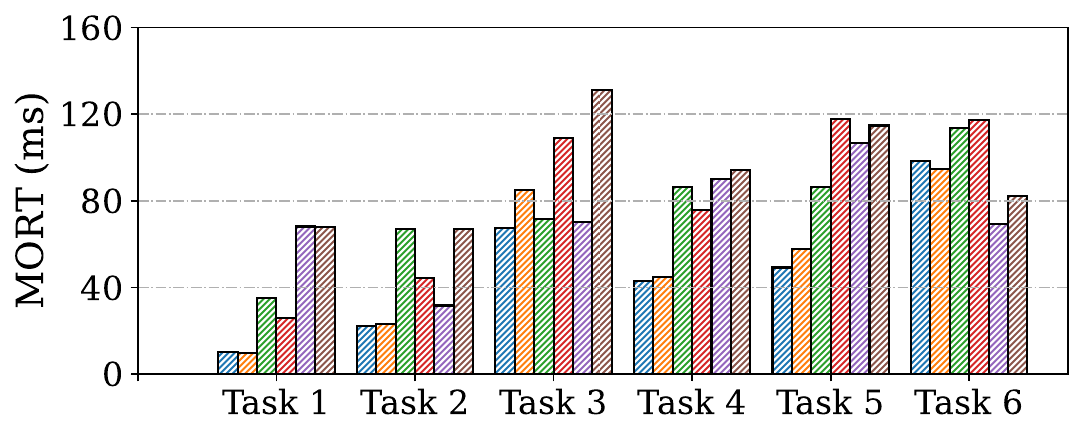}
    \vspace{-1.5\baselineskip}\caption{Jetson Xavier NX}
    \label{fig:overall_mort_nx}
    \end{subfigure}\hfill
    \begin{subfigure}[b]{0.49\linewidth}
        \vspace{-0.5\baselineskip}
        \includegraphics[width=\linewidth]{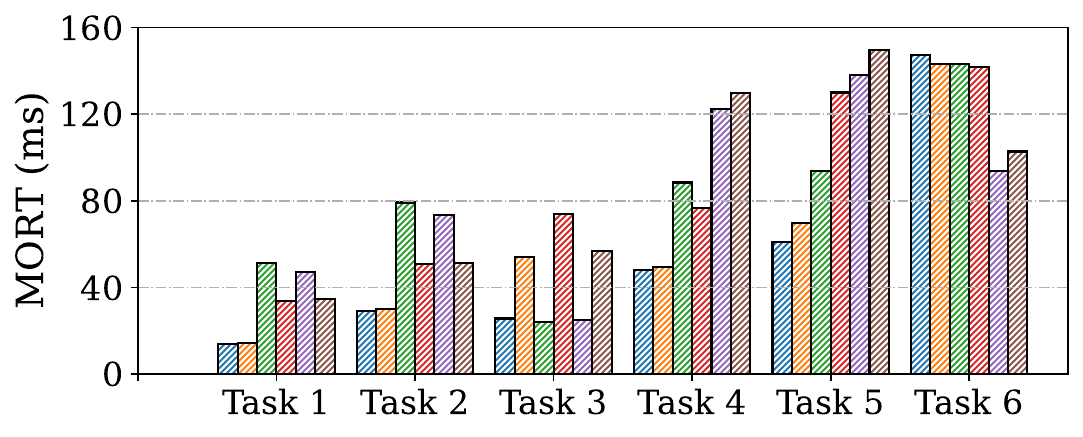}
        \vspace{-1.5\baselineskip}\caption{Jetson Orin Nano}
        \label{fig:overall_mort_orin}
    \end{subfigure}
\caption{Maximum observed response time on two platforms}
\label{fig:overall_mort}
\end{figure}

\begin{figure}[t]
\vspace{-10pt}
    \centering
    \includegraphics[width=\linewidth]{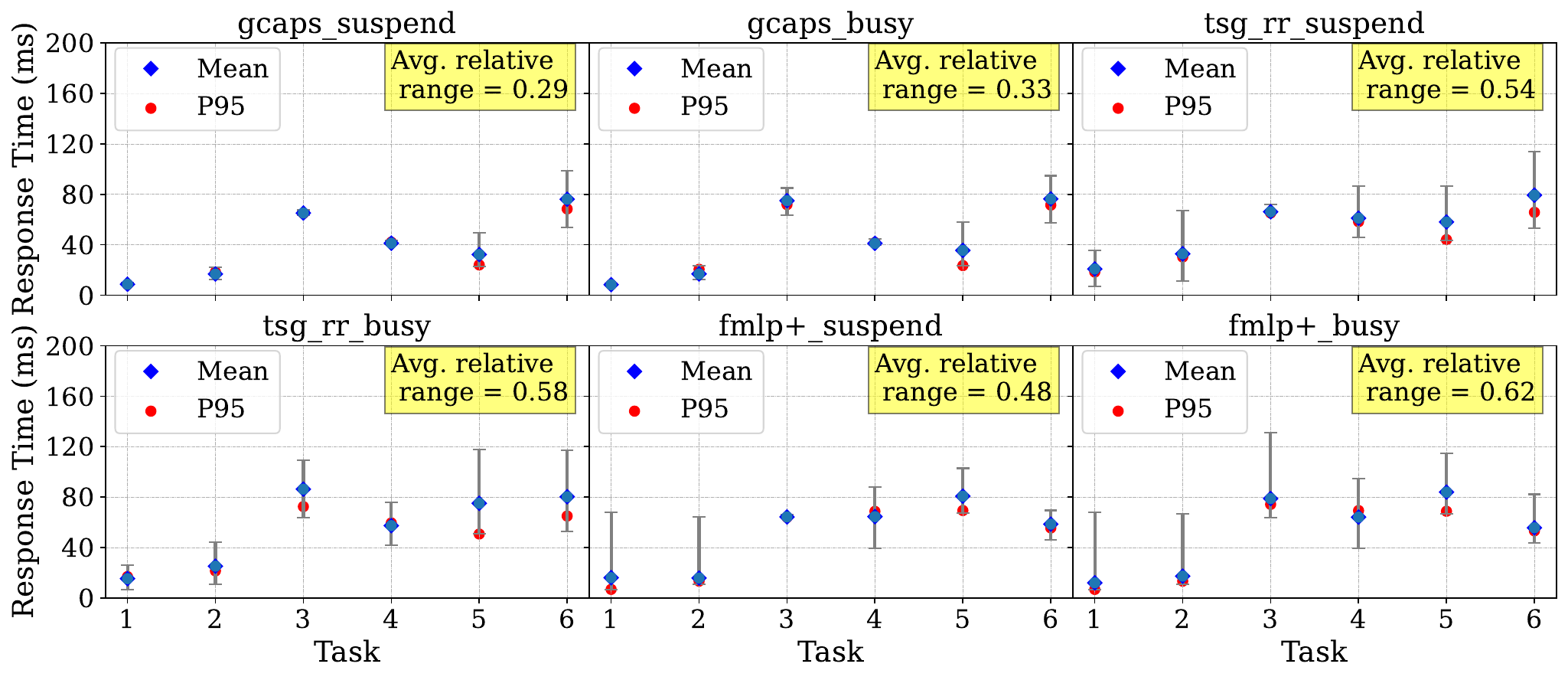}
\vspace{-1.5\baselineskip}     
\caption{Observed response time variations on Jetson Xavier}
\label{fig:individual_mort_nx}
\end{figure}

\noindent\textbf{Case Study.} 
We conducted a case study on the aforementioned platforms to evaluate the performance and effectiveness of the proposed preemptive GPU scheduling mechanism. Table~\ref{tab:taskset_case_study_nx} provides a summary of the taskset employed in this study, and we show the tasks' WCET collected on Jetson Xavier NX. The benchmarks are from Nvidia CUDA Samples~\cite{cuda_sample}. The execution requirements of the tasks ($C_i$ and $G_i$) are obtained by profiling the benchmarks and rounding them up. $T_i$ ($= D_i$) is chosen based on $C_i$ and $G_i$ to ensure task utilization falls between 0.05 and 0.35. The CPU assignment is based on the consideration of load balancing.

The tasks in the table are arranged in descending order of priority, and each task's GPU segments use the same OS-level priority as its CPU segments. Tasks 3 is a CPU-only task with $G_i = 0$, while the remaining tasks involve GPU computations. Tasks 6 and 7 are categorized as best-effort tasks, as they are not assigned real-time priority. Task 7 is a graphic application running at 16 FPS to stress the GPU.
To suspend a task during its GPU execution, we used CUDA events with the  \texttt{cudaEventBlockingSync} flag. We compared our approaches against \texttt{tsg\_rr} (default round-robin scheduling in Nvidia GPU driver) and \texttt{fmlp+} (synchronization-based approach).

We released the tasks at the same time and executed them for a duration of 30s during which we measured the maximum observed response time (MORT) for each real-time task. The results are depicted in Fig.~\ref{fig:overall_mort_nx}. 
Generally, the proposed approach can provide low MORT for real-time tasks, particularly for higher-priority tasks such as Task 1 and Task 2. This indicates that the GCAPS approach prioritizes these tasks effectively.
However, it is apparent that the best-effort tasks, such as Task 6, experience a trade-off, displaying higher MORT under the proposed approaches than \texttt{fmlp+}. This is also due to the reason that the low-priority tasks can benefit from the blocking effect under \texttt{fmlp+} approaches. We omit the results for task 7 since it is a graphic application and its performance is measured through FPS. According to our observation, under each scheduling approach, an average FPS of around 15 can be maintained.

Fig.~\ref{fig:individual_mort_nx} illustrates the observed response time of each task, with error bars representing the deviation from the mean; above for "Max-Mean" and below for "Mean-Min" respectively. The "Average relative range" is calculated as "(Max-Min)/Max" to reflect variability.
Both \texttt{gcaps\_suspend} and \texttt{gcaps\_busy} show a tendency to more consistent response times for real-time higher-priority tasks, as evidenced by more compact error bars in Fig.~\ref{fig:individual_mort_nx} and small variability 
compared to \texttt{tsg\_rr} and \texttt{fmlp+}.
\texttt{fmlp+} exhibits higher variability in the observed response time primarily due to blocking which has significantly increased the response time for the real-time tasks. 
Meanwhile, \texttt{tsg\_rr} displays medium-sized error bars across the tasks, due to the use of a fairer allocation of GPU resources but without introducing blocking.


\begin{table}[]
\small
\begin{tabular}{r|rr|rr|rr|rr}
\hline
\multicolumn{1}{l|}{\multirow{2}{*}{Task}} & \multicolumn{2}{l|}{tsg\_rr\_suspend}                & \multicolumn{2}{l|}{tsg\_rr\_busy}                     & \multicolumn{2}{l|}{gcaps\_suspend}                  & \multicolumn{2}{l}{gcaps\_busy}                     \\ \cline{2-9} 
\multicolumn{1}{l|}{}                      & \multicolumn{1}{l}{MORT} & \multicolumn{1}{l|}{WCRT} & \multicolumn{1}{l}{MORT} & \multicolumn{1}{l|}{WCRT}   & \multicolumn{1}{l}{MORT} & \multicolumn{1}{l|}{WCRT} & \multicolumn{1}{l}{MORT} & \multicolumn{1}{l}{WCRT} \\ \hline
1   & 45.33    & 60      & 26.13     & 60        & 10.15     & 16    & 9.68      & 16                       \\
2   & 66.97    & 73.6      & 44.47     & 73.6      & 22.36     & 32    & 23.28     &32                       \\
3   & 71.84    & 76         & 109.14    & 129.2    & 67.39     & 75    & 85.01     & 111                      \\
4   & 86.50    & 98.2     & 75.64     & 192.2   & 43.17   & 59    & 44.91     & 59                       \\
5   & 86.62    & 127.8     & 117.68    & \red{Failed} & 49.24  & 79    & 57.93     & 79                       \\ \hline
\end{tabular}
\caption{Comparison of MORT (ms) and WCRT (ms) on Jetson Xavier}
\label{tab:mort_wcrt_nx}
\end{table}

Table~\ref{tab:mort_wcrt_nx} lists the comparison of MORT and the worst-case response time (WCRT) bounds computed using our analysis given in Sec.~\ref{sec:analysis} for the default round-robin scheduling and our proposed approaches. \texttt{tsg\_rr\_busy} failed the response time test at Task 5 due to the conservative nature of the analysis for busy-waiting tasks. The results of \texttt{fmlp+} are omitted since the tests failed at Task 1, while we observed no deadline misses for this taskset running on the real systems. This proves that our proposed preemptive GPU scheduling approach can offer tighter WCRT bounds for higher-priority tasks.

We run the same experiments on Nvidia Jetson Orin Nano, an embedded GPU platform with the latest Ampere architecture, and the similar trends of MORT are shown in Fig.~\ref{fig:overall_mort_orin}.


\begin{figure}[h!]
    \vspace{-10pt}
    \centering
    \begin{subfigure}[b]{0.48\linewidth}
        \includegraphics[width=\linewidth]{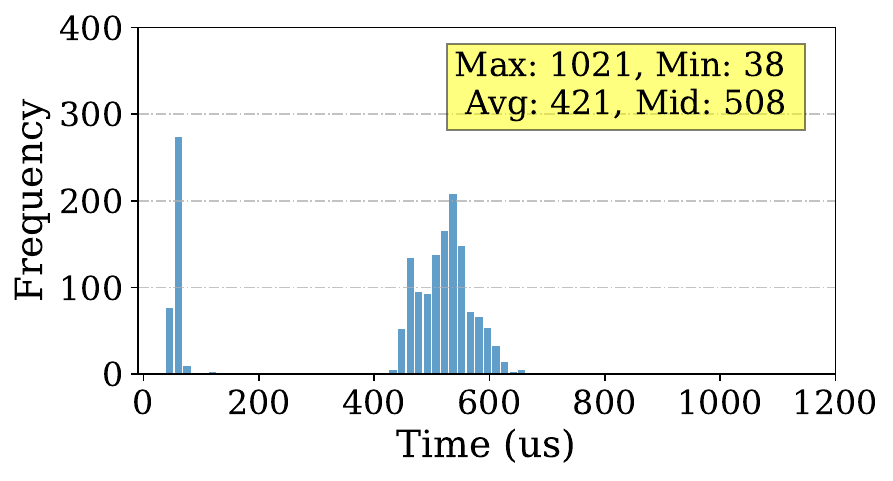}
    \vspace{-1.5\baselineskip}\caption{Jetson Xavier NX}
    \end{subfigure}
    \begin{subfigure}[b]{0.48\linewidth}
        \includegraphics[width=\linewidth]{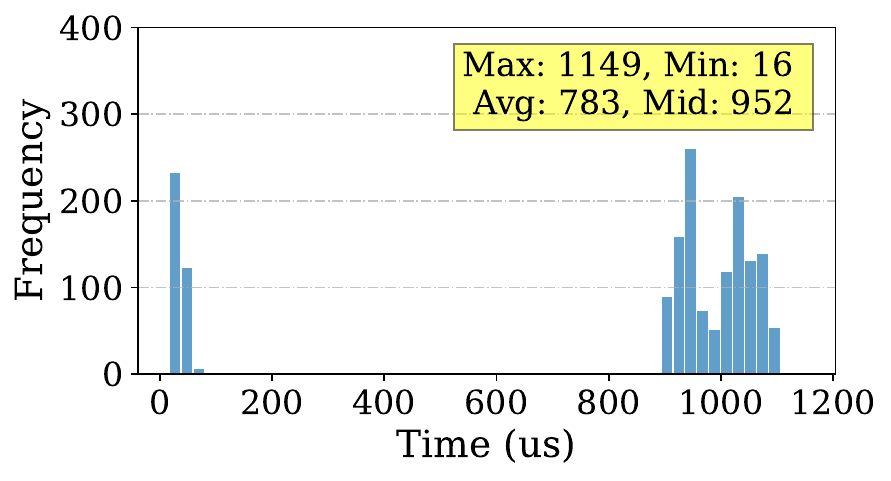}
        \vspace{-1.5\baselineskip}\caption{Jetson Orin Nano}
    \end{subfigure}
\caption{Histogram of runlist update overhead}
\label{fig:overhead_dist}
\end{figure}

\noindent\textbf{Runlist Update Overhead. } We also measured the overhead of runlist update, $\epsilon$ (Def.~\ref{def:runlist_update_delay}), while running the taskset in the case study. The data and the distribution is shown in Fig.~\ref{fig:overhead_dist}. 
The lower mode in the distribution indicates requests that do not necessarily require runlist updates, and it mainly includes the cost of accessing the IOCTL system call. In our experimental settings, these two boards have similar CPU frequencies at about 1.5GHz while Jetson Xavier NX has a much higher GPU frequency of 1.1GHz than Jetson Orin Nano's 625MHz.
Both platforms exhibit a maximum overhead of about 1 ms, which is higher than the range reported in prior work~\cite{capodieci2018deadline}. We suspect this is due to the relatively lower frequency of our GPUs and it could be optimized in future generations of GPU architectures, as can be seen with Orin's case (10\% higher overhead despite half the frequency). Nonetheless, we consider the cost acceptable based on our schedulability experiments conducted with a similar overhead. 

\begin{figure}[h!]
    \vspace{-10pt}
    \centering
    \includegraphics[width=0.4\linewidth]{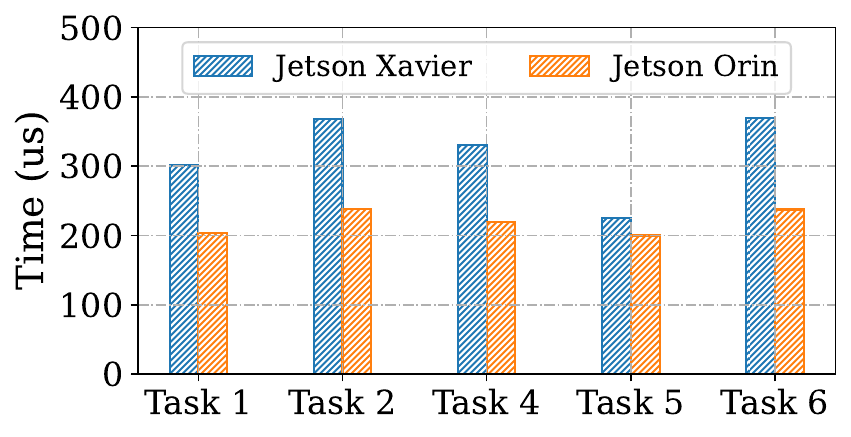}
\caption{Avg. Overhead per TSG Context Switching}
\label{fig:tsg_ctxsw_time}
\end{figure}

\noindent\textbf{TSG Context Switching Overhead. }
We designed a separate experiment to measure the TSG context switching overhead, $\theta$ (defined in Sec.~\ref{sec:background}). The main idea is to use Eq.~\eqref{eq:gpu_rr}. To accurately gauge the number of TSGs, we incorporated a dummy loop within each kernel to extend the kernel duration so that it ensures that multiple timeslices are required to complete the execution. We launched different numbers of instances simultaneously of each workload given in Table.~\ref{tab:taskset_case_study_nx}. In an ideal scenario, running $\nu$ identical kernels concurrently would lead to a slowdown factor of $\nu$, and we can estimate the TSG context switching overhead based on the difference between the ideal and the actual slowdown. Specifically, we first measured the independent completion time of a kernel as $E_{1}$. We then launched $\nu$ identical kernels simultaneously and recorded the completion time as $E_{\nu}$. This approach allows us to compute the TSG switching overhead as follows:
\begin{equation}
\small
\begin{aligned}
    \theta = \frac{E_{\nu} - \nu \cdot E_{1}}{\nu \cdot E_{1}} \cdot L
\end{aligned}
\end{equation}
where $L$ is set to 1000$\mu$s in the default Tegra driver. The task slowdown and the estimated TSG context switching times are listed in Fig.~\ref{fig:tsg_ctxsw_time}, where both platforms demonstrate an average TSG context switching overhead greater than 200$\mu$s which is not trivial, especially for long-running kernels. It is worth noting that the overhead on Jetson Orin is lower than that one Jetson Xavier, which is opposite to the case of runlist update delay ($\epsilon$). This might be attributed to several factors related to architectural differences that Jetson Orin has a better pipeline management and a more efficient context-saving mechanism.

\vspace{-10pt}
\section{Conclusion}
\nocite{wang2023dissertation} \nocite{wang2024unleashing} 
In this paper, we present a preemptive priority-based scheduling approach for GPU-using tasks in a multi-core real-time system equipped with an Nvidia GPU. 
We first discussed how the Nvidia Tegra GPU driver works and presented the design of GCAPS, our priority-based preemptive GPU context scheduling approach. 
Then, we provided a comprehensive response time analysis for both the default round-robin scheduling of the device driver and our proposed approach. To the best of our knowledge, this was the first attempt to formally analyze the worst-case response time under the default GPU driver's time-shared GPU context scheduling mechanism.
Through empirical evaluations, we have demonstrated the effectiveness of our approach in enhancing schedulability compared to synchronization-based approaches and the default driver. Additionally, our case study shows the benefits of our approach in predictability and responsiveness over the default GPU driver and prior work. 

Future work can focus on further optimizing and refining the proposed approach and exploring additional scheduling strategies such as dynamic priority. Combining our device-driver level approach with GPU partitioning mechanisms will also be an interesting direction.

\small
\bibliography{ref.bib}

\end{document}